\documentclass[11pt, JPCM]{iopart}

\usepackage{graphicx}
\usepackage{wrapfig}
\usepackage{subfig}
\usepackage{sidecap}
\expandafter\let\csname equation*\endcsname\relax
\expandafter\let\csname endequation*\endcsname\relax
\usepackage{amssymb}
\usepackage{bigints}

\begin{document} 

\title[Packing in Protein Cores]{Packing in Protein Cores}
\author{J. C. Gaines$^{1,2}$, A. H. Clark$^{3}$,  L. Regan$^{2,4,5}$ and C. S. O'Hern$^{3,1,2,6,7}$} 
\address{$^1$ Program in Computational Biology and Bioinformatics, Yale University, New Haven, Connecticut, 06520}
\address{$^2$ Integrated Graduate Program in Physical and Engineering Biology (IGPPEB), Yale University, New Haven, Connecticut, 06520}
\address{$^3$ Department of Mechanical Engineering \& Materials Science,
Yale University, New Haven, Connecticut, 06520}
\address{$^4$ Department of Molecular Biophysics \& Biochemistry, Yale University, New Haven, Connecticut, 06520}
\address{$^5$ Department of Chemistry, Yale University, New Haven, Connecticut, 06520}
\address{$^6$ Department of Physics, Yale University, New Haven, Connecticut, 06520}
\address{$^7$ Department of Applied Physics, Yale University, New Haven, Connecticut, 06520}

\begin{abstract}
Proteins are biological polymers that underlie all cellular
functions. The first high-resolution protein structures were
determined by x-ray crystallography in the 1960s. Since then, there
has been continued interest in understanding and predicting protein
structure and stability. It is well-established that a large
contribution to protein stability originates from the sequestration
from solvent of hydrophobic residues in the protein core. How are such
hydrophobic residues arranged in the core? And how can one best model
the packing of these residues? Here we show that to properly model the
packing of residues in protein cores it is essential that amino acids
are represented by appropriately calibrated atom sizes, and that
hydrogen atoms are explicitly included. We show that protein cores
possess a packing fraction of $\phi \approx 0.56$, which is
significantly less than the typically quoted value of $0.74$ obtained
using the extended atom representation.  We also compare the results
for the packing of amino acids in protein cores to results obtained
for jammed packings from disrete element simulations composed of
spheres, elongated particles, and particles with bumpy surfaces.  We
show that amino acids in protein cores pack as densely as disordered
jammed packings of particles with similar values for the aspect ratio
and bumpiness as found for amino acids.  Knowing the structural
properties of protein cores is of both fundamental and practical
importance. Practically, it enables the assessment of changes in the
structure and stability of proteins arising from amino acid mutations
(such as those identified as a result of the massive human genome
sequencing efforts) and the design of new folded, stable proteins and
protein-protein interactions with tunable specificity and affinity.


\end{abstract}

\clearpage	
\section{Introduction}

Proteins are biological polymers that play important roles in cellular
processes ranging from the purely structural to the actively
catalytic. Proteins are linear chains of different combinations of the
$20$ naturally occurring amino acid residues with variable chain
lengths from tens to tens of thousands. A key feature that
distinguishes proteins from other polymers is that each folds into a
unique three-dimensional structure. Proteins typically fold
spontaneously in aqueous solution at room temperature. The amino acid
sequence is the only information required to specify a protein's
unique structure \cite{Dill_1990,Rose_2006}.

The amino acids can be grouped into two main categories: hydrophobic
and hydrophilic. Hydrophobic residues form the solvent-inaccessible
core of a protein and hydrophilic residues, both polar and charged,
are on the solvent-accessible surface. As of 2017, the structures of
more than 125,000 proteins have been determined, primarily by x-ray
crystallography, with a median resolution of $\approx 2.5$~\AA~and
deposited in the protein data bank (PDB)~\cite{Berman_2000}. This large
database of atomic coordinates provides a wealth of structural information
that can be used to analyze the physical properties of proteins and to understand how proteins interact and carry out
their functions  \cite{Dunbrack_1997,LoConte_1999,Glaser_2001,Keskin_2004,Bordner_2005,Reichmann_2007,Sheffler_2009, London_2010,HS_Z2011,HS_Z2013, Gaines_2016}.

 \begin{SCfigure}[][!b]
\includegraphics[width=3in]{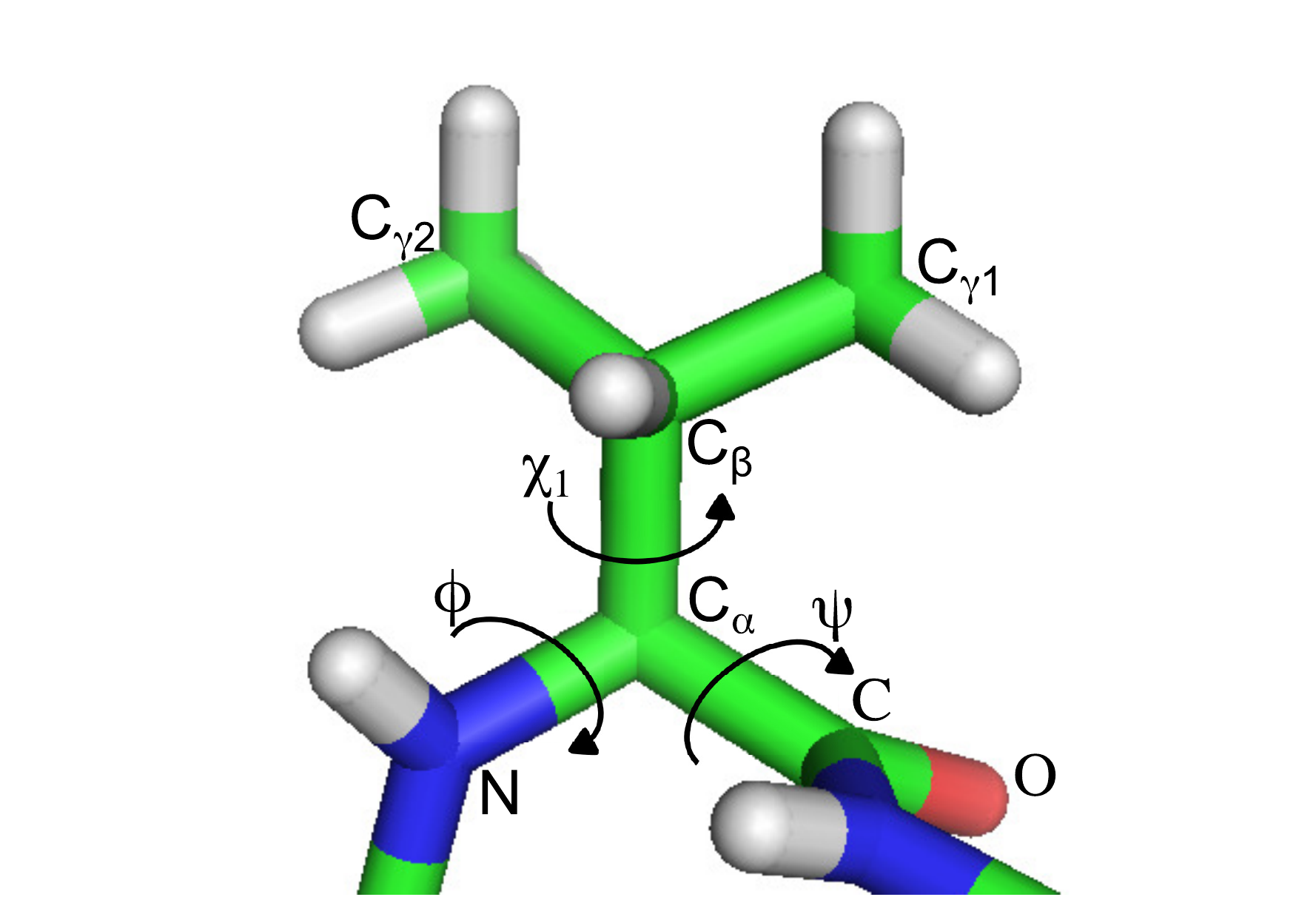}
\caption{Stick representation of a valine (Val) residue with each atom shown in
a different color: C (green), N (blue), O (red), and H
(white). The heavy (non-hydrogen) atoms are also labeled. The two backbone dihedral
angles $\phi$ and $\psi$  and one side chain dihedral angle $\chi_1$ 
(defined by the atoms N-C$_\alpha$-C$_\beta$-C$_{\gamma1}$) are indicated.}
\label{Stick_Val}
\end{SCfigure}

Each amino acid is made up of the same backbone unit of four heavy
(non-hydrogen) atoms, N-C$_\alpha$-C-O, and different combinations of
side chain atoms that branch from the C$_\alpha$ atom
(Fig.~\ref{Stick_Val}). The repeating units are joined by a peptide
bond between the carboxyl carbon (C) of a given amino acid and the
nitrogen (N) of the next. All bond lengths and bond angles are
specified by the same basic stereochemistry that defines the
structures of small molecules \cite{Engh_1991,Allen_2002}.  The
three-dimensional structure that a protein adopts is specified by the
amino acid dihedral angles. For each amino acid in the protein chain,
there are two backbone dihedral angle degrees of freedom, $\phi$ and
$\psi$, and $N_s$ side chain dihedral angle degrees of freedom,
$\chi_1, \ldots, \chi_{N_s}$. (See Fig.~\ref{Stick_Val}.) $N_s$ ranges
from zero (for alanine and glycine) to five (for arginine). The third
backbone dihedral angle is typically constrained to be $\omega =
180^{\circ}$ or $0^{\circ}$.  Repetition of certain backbone $\phi$
and $\psi$ values in a stretch of amino acids gives rise to specific
secondary structures, such as $\alpha$-helices and $\beta$-sheets
\cite{Ramakrishnan_1963, Ramakrishnan_1965}. All proteins are formed
from different combinations of $\alpha$-helix, $\beta$-sheet, and
`random coil' structures.  Interactions between different elements of
secondary structure are stabilized by interactions between the side
chains \cite{Bryson_1995, Munson_1996, Smith_1995}. In addition, side
chain interactions on the surfaces of proteins also specify how
different proteins bind to each other and to other molecules
\cite{Glaser_2001}.

\begin{SCfigure}
\subfloat{\includegraphics[height = 2in]{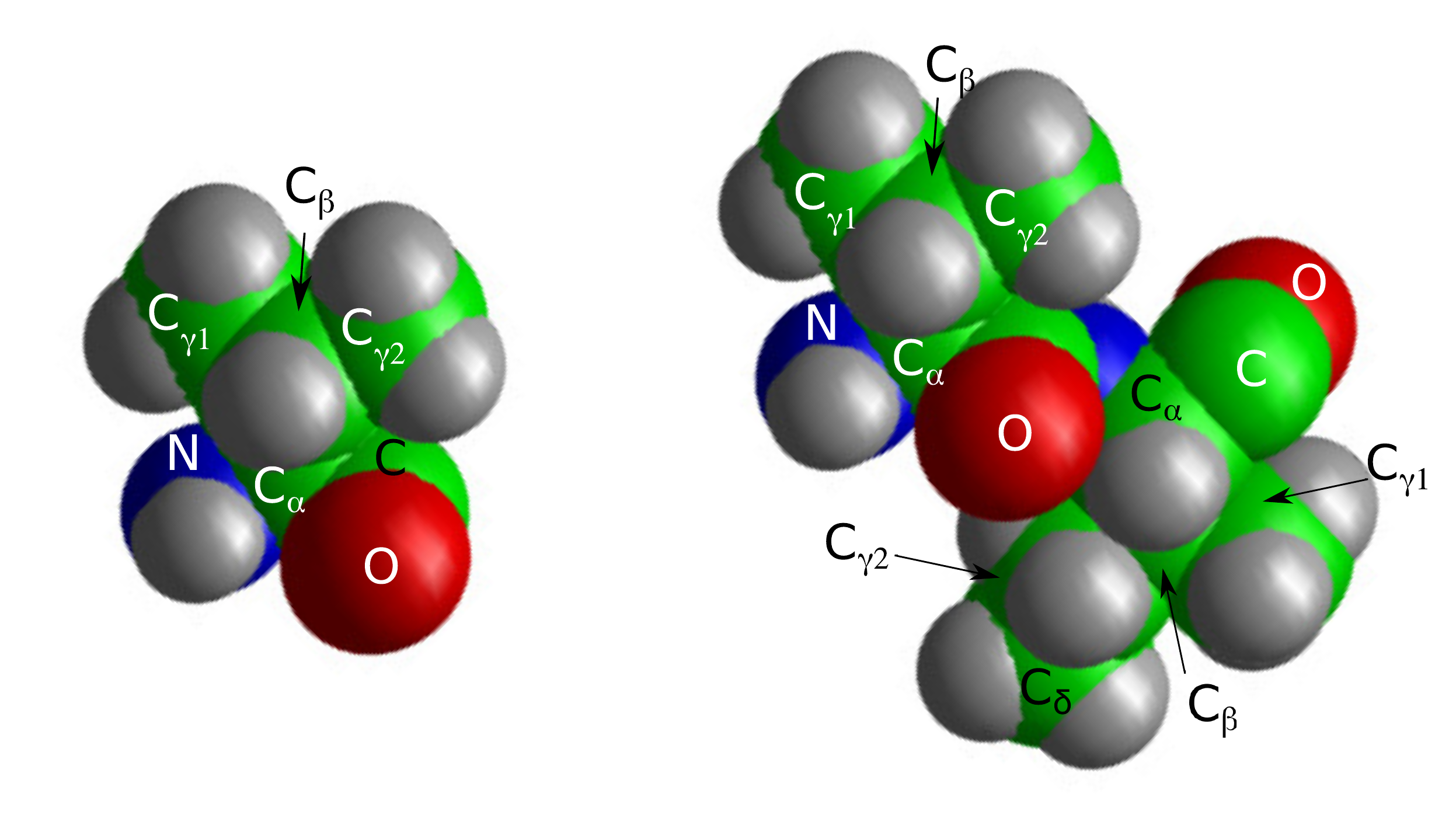}}
\caption{(left) Illustration of a Val residue with each atom represented as a
sphere: C (green), O (red), N (blue), and H (grey). (right)
Val and Ile residues with connected backbones taken from PDB 1K5C. In 
both panels, heavy atoms are labeled.}
\label{sphere_residues}
\end{SCfigure}

A minimal physical model for an amino acid is a composite particle 
formed from connected spheres with stereochemical constraints
(Fig. \ref{sphere_residues}). As is clear from Fig.
\ref{sphere_residues}, amino acids are non-spherical objects with
complex shapes. Thus, we can imagine proteins as strings of
interconnected non-spherical objects that fold into compact three-dimensional
structures. Many prior studies have argued that the cores of folded
proteins are tightly packed. For example, several studies have
measured the ratio between the volume of a core amino acid and its Voronoi
volume to be greater than 70\%, which suggests dense crystalline packing
\cite{Richards_1974, Liang_2001}. In addition, experimental studies find that mutations in protein cores from small to large residues typically destabilize the protein, suggesting that
there is very little empty space present to accommodate additional
atoms \cite{Lim_1989, Lim_1991}.  

In this review, we summarize prior work on the structural properties
of protein cores and provide strong evidence that although protein
cores are densely packed, they are not as densely packed as
crystalline solids.  Instead, protein cores possess packing fractions
of $\sim 0.56$~\cite{Gaines_2016}. Even though this value is lower
than that for crystalline solids ({\it e.g.} $0.74$ for
face-centered-cubic crystals), protein cores are solid-like with very
little free volume that would allow side chain motion. We also show
that static packings of particles with complex, non-spherical shapes possess
packing fractions below $0.6$, yet still display solid-like properties
and that the amino acids in protein cores can be modeled as random,
densely packed non-spherical objects.  We then relate our computational 
studies of dense packing in protein cores to experimental studies of 
mutations that are able to alter the structure and stability of proteins.  
  
\section{Packing efficiency in protein cores}
\label{phi_section}

By determining the packing fraction of protein cores one can begin to
understand their structural and mechanical properties.  For example,
the shear modulus ({\it i.e.} the material response to applied shear
stress) typically increases monotonically with the packing fraction
since the number of stress-bearing interatomic contacts increases with
the packing fraction \cite{Ohern_2003}. Thus, the rigidity of proteins
is strongly correlated with the packing density
\cite{Gekko_2015,Chalikian_1995}. In addition, knowing the packing
density is vital for predicting changes in stability from mutations to
protein cores, many of which are disease-associated
\cite{Gao_2015}. Accurate calculations of the packing density are also
necessary to predict structure from sequence and to design new stable
proteins \cite{Sheffler_2009, Regan_2015,Sheffler_2010}.

\begin{figure}[t]
\subfloat{\includegraphics[trim=1in 2.7in 1.75in 2.85in,clip=true, totalheight =1.95in]{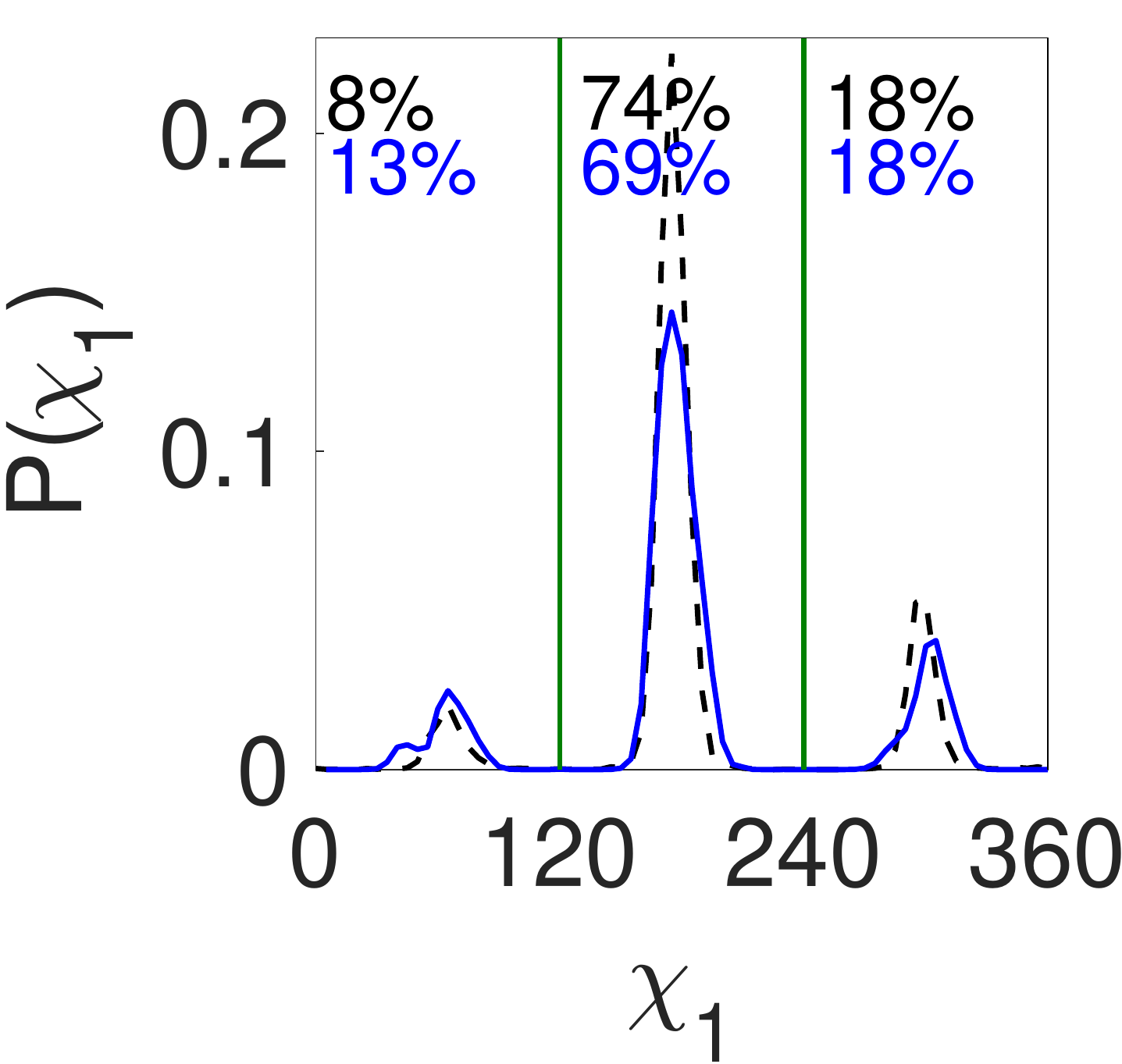}}
\subfloat{\includegraphics[trim=.5in 2.7in 2in 2.85in,clip=true,  totalheight =1.95in]{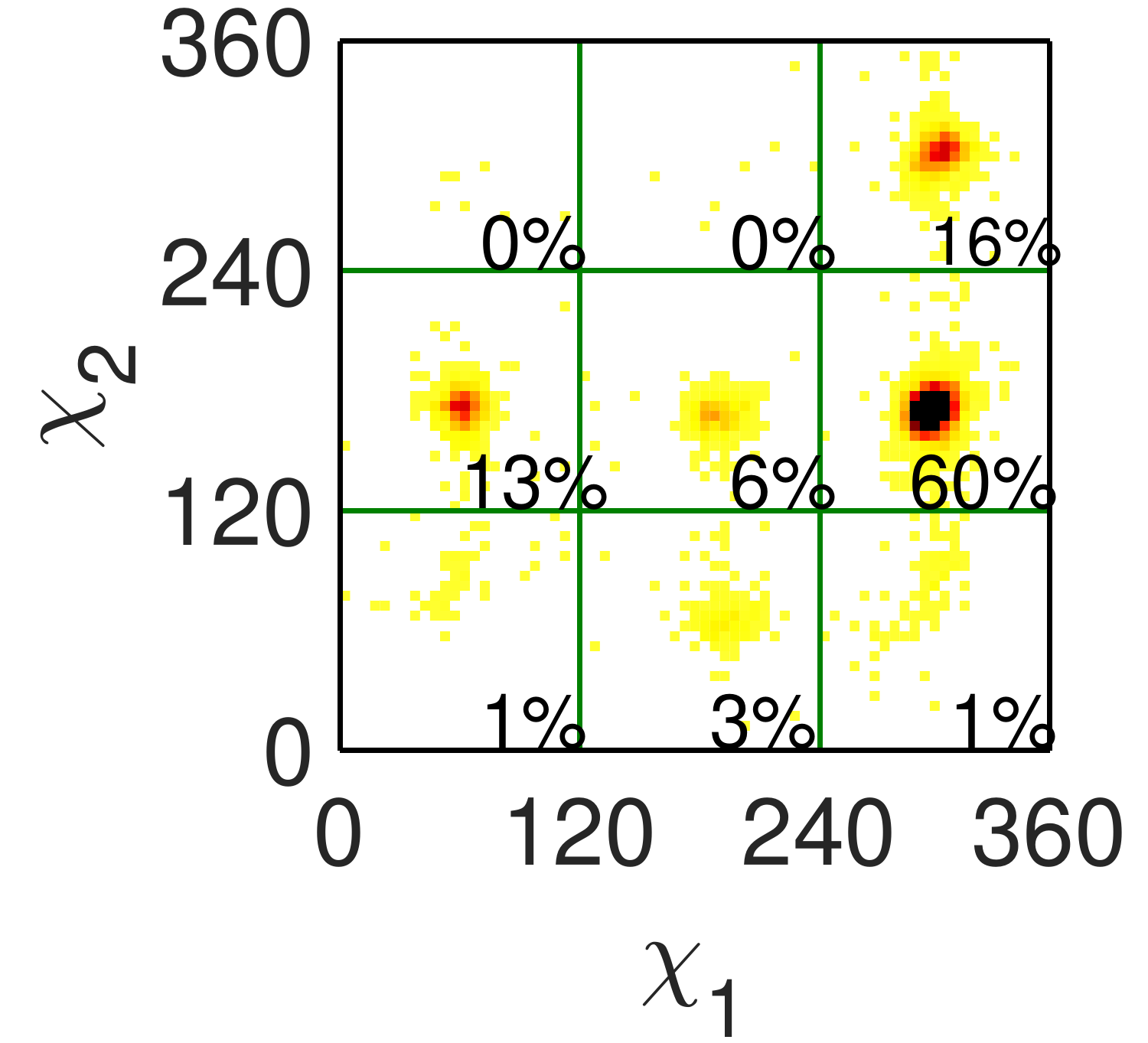}}
\subfloat{\includegraphics[trim=.5in 2.7in 0.5in 2.85in,clip=true,  totalheight =1.95in]{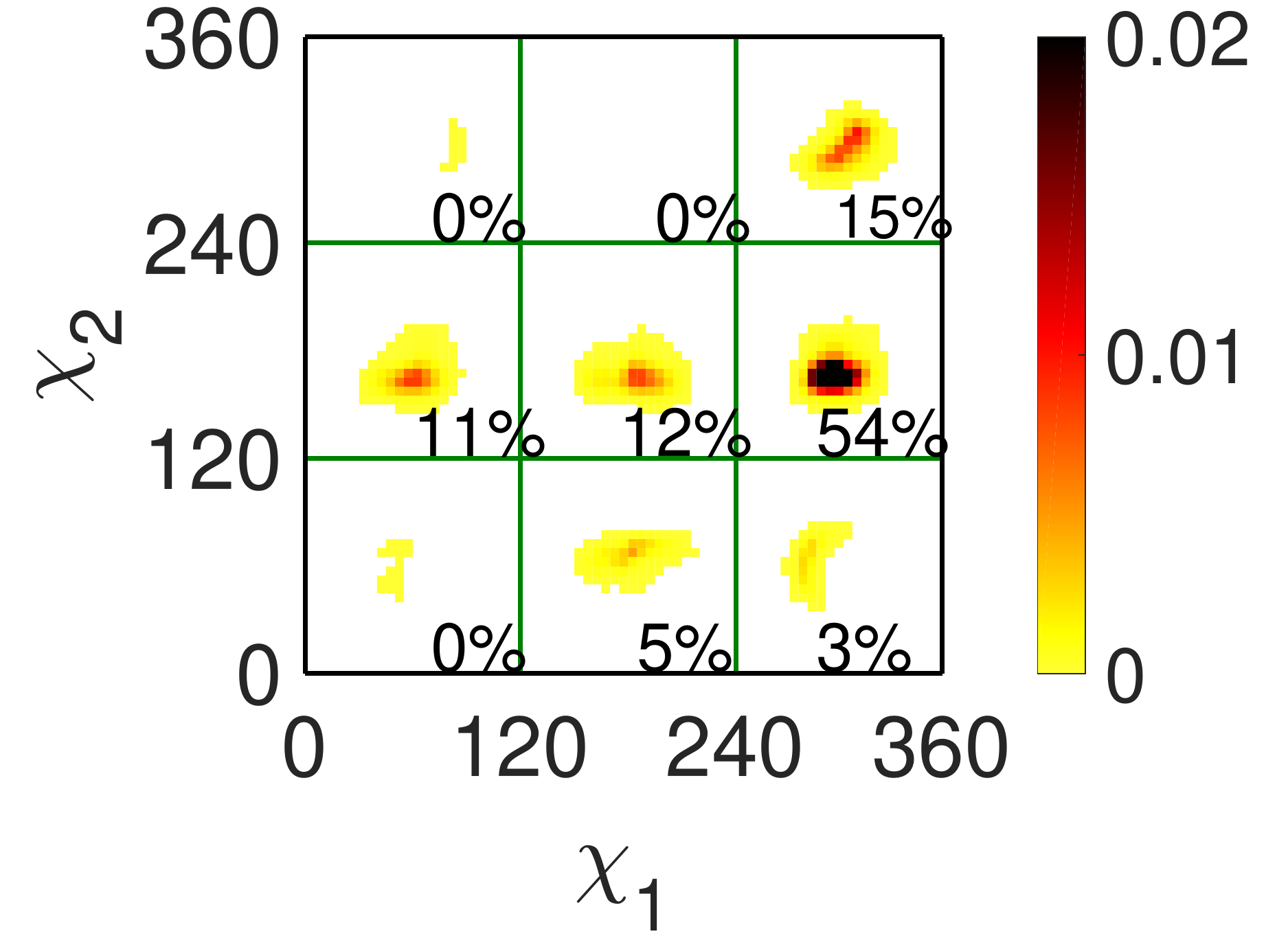}}
\caption{(left) The observed side chain dihedral angle probability
distribution (black dotted line), $P(\chi_1)$, for Val residues in a
database of high resolution protein crystal structures (described in
\cite{Gaines_2016,Wang_2003, Wang_2005}) compared to $P(\chi_1)$
predicted by the hard-sphere dipeptide mimetic model for Val using
the explicit hydrogen atom representation (blue solid
line). (center) The observed side chain dihedral angle probability
distribution $P(\chi_1, \chi_2)$ for Ile. (right) The predicted side
chain dihedral angle distribution for Ile using the hard-sphere model. The probabilities
increase from light to dark. The percentages give the fractional
probabilities that occur in each of the three and nine rotamer bins in 
the left panel and center/right panels, respectively. The center 
and right panels are reprinted  with permission from 
[J. C. Gaines, W. W. Smith, L. Regan, and C. S. O'Hern, Phys. Rev. E, 
93, 032415, 2016.] Copyright (2016) by the American Physical Society.}
\label{Dihedral_dist}
\end{figure}

One of the first studies of the packing density of protein cores was
performed by Richards in 1974. At this time, only a few protein
crystal structures were available. Richards focused on two proteins:
lysozyme and ribonuclease S~\cite{Richards_1974}. When a protein
structure is obtained from x-ray crystallography, the resolution of
the structure typically does not allow for the placement of the
hydrogen atoms in the protein. In the past, researchers circumvented
this problem by implementing an ``extended atom" model, where the
atomic radii of each heavy atom are increased by a factor that depends
on the number of hydrogen atoms that are bound to
it~\cite{Richards_1974, Liang_2001,Tsai_1999}. New computational
techniques allow for the accurate placement of hydrogen atoms in a
protein crystal structure~\cite{Word_1999, Word_1999b}, which provides
a more detailed ``explicit hydrogen" model of proteins. Since hydrogen atoms
comprise $\sim 50\%$ of the atoms in a protein, the extended atom approximation can have major
effects on the accuracy of the structural model of the protein.

To accurately assess the packing fraction of proteins, one must
calibrate and select proper atomic radii. In our recent
work~\cite{Gaines_2016}, we have chosen atomic radii that when used in
a hard-sphere model of a dipeptide mimetic can reproduce the observed
side chain dihedral angle distributions of non-polar amino acids in a
database of high resolution crystal structures~\cite{Gaines_2016,Wang_2003, Wang_2005, HS_Z2012,
  HS_Z2014}. The values for the seven
atomic radii are $C_{sp^3}$, $C_{\rm aromatic}$: $1.5$ {\AA}; $C_O$:
$1.3$ {\AA}; $O$: $1.4$ {\AA}; $N$: $1.3$ {\AA}; $H$: $1.10$ {\AA};
and $S$: $1.75$ {\AA}. In Fig. \ref{Dihedral_dist}, we show that the
side chain dihedral angle distributions predicted using the
hard-sphere model for a Val and Ile dipeptide agree with the observed
side chain dihedral angle distributions. We have shown similar
agreement between the observed and predicted side chain dihedral angle
distributions for Cys, Leu, Met, Phe, Thr, Trp, Tyr, and Ser
\cite{HS_Z2014}.  The atomic radii are similar to values of van der
Waals radii reported in other studies, and typically smaller than those used in
extended atom models (Table \ref{radii}) \cite{ Ramakrishnan_1965, Richards_1974,Tsai_1999, HS_Z2012,
 Bondi_1964, CDC, Seeliger_2007, Pauling_1948,
  Porter_2011,  Chothia_1975,  Li_1998,
  Momany_1974, Allinger_1980}.

\begin{table}
\begin{center}
\begin{tabular}{|c|c|c|c|c|}
	\hline
	Atom Type & Hard-sphere Model &Word 1999 \cite{Word_1999} & Richards 1974 \cite{Richards_1974} & Liang 2001 \cite{Liang_2001} \\ \hline
	C$_{sp^3}$ & 1.5 & 1.65  & 2.0 & 1.88\\
	C$_{aromatic}$ &1.5 & 1.65 & 1.7 & 1.61 - 1.76 \\
	C$_{O}$ & 1.3 & 1.65 & 1.7 & 1.61 - 1.76\\
	N & 1.3 & 1.55 & 1.7 & 1.64\\
	O & 1.4 & 1.4 & 1.4 - 1.6 & 1.42 - 1.46\\
	S & 1.75 & 1.8 & 1.8 & 1.77\\
	H & 1.1 & 1.17 & N/A  & N/A\\
	\hline
	
\end{tabular}
\caption{Atomic radii used in the hard-sphere model and 
three other studies (one using explicit hydrogens~\cite{Word_1999} and 
two others using 
the extended atom model~\cite{Richards_1974,Liang_2001}). All values are given in {\AA}.}
\label{radii}
\end{center}
\end{table}

The packing fraction of each residue in a protein core can be
calculated using
\begin{equation}
\phi_r = \frac{\sum_i \mathrm{V}_{i}}{\sum_i \mathrm{V}^v_{i}},
\label{phi_eq}
\end{equation}
where $\mathrm{V}_{i}$ is the `non-overlapping' volume of atom {\it
  i}, $\mathrm{V}^v_{i}$ is the Voronoi volume of atom {\it i}, and
the summations are over all atoms of a particular residue. We also
calculate the packing fraction of a protein core, $\phi_c$, where both
summations are over all atoms of all residues in a particular protein
core.  Voronoi cells were obtained for each atom using Laguerre
tessellation, where the placement of the Voronoi cell wall is based on
the relative radii of neighboring atoms (which is the same as the
location of the plane that separates overlapping
atoms)~\cite{Gaines_2016,Rycroft_2009}.  $V_i$ was calculated by
splitting overlapping atoms by the plane of intersection between the
two atoms. 

Our analysis focuses on residues in protein cores. We have identified
all core residues in a database of high resolution crystal structures
(described in \cite{Gaines_2016,Wang_2003, Wang_2005}) using a method
described previously \cite{Gaines_2016,Caballero_2016}. In brief,
non-core atoms are identified as those that are on the surface of the
protein or near an interior void with a radius $\geq
1.4$~$\mathrm{\AA}$. In this strict definition, a core residue is
defined as any residue containing exclusively core atoms (including
hydrogen atoms).  This method identifies atoms adjacent to voids in
the protein and removes them from the calculation of the packing
fraction. According to this definition and using the explicit hydrogen
representation, the proteins we considered have an average of $15$
core residues of which $80\%$ are Ala, Cys, Gly, Ile, Leu, Met, Phe,
and Val.

\begin{figure}
\includegraphics[width=6.5in]{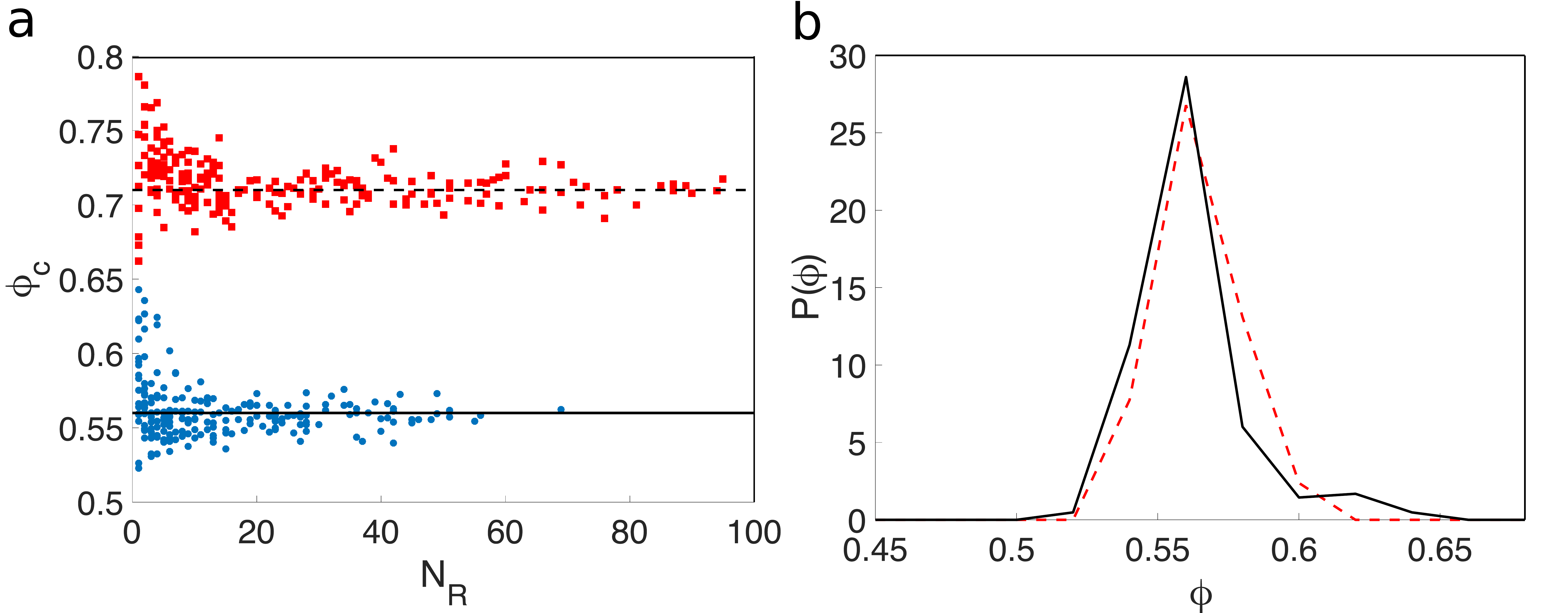}
\vspace{-.15in}
\caption{(a) A comparison of the packing fraction $\phi_c$ of protein cores as a function of the number
of core residues ($N_R$) using the explicit hydrogen (blue circles)
and extended atom (red squares) representations. More
residues are designated as core using the extend atom model
(25 on average) than using the explicit hydrogen model (15
on average). The dashed and solid horizontal lines indicate the average packing fraction of each system, $\phi_c = 0.71$ for extended atom and $\phi_c = 0.56$ for explicit hydrogen. (b) The probability distribution (red dotted line) of packing
fractions at jamming onset $P(\phi)$ from simulations of mixtures of
individual residues found in protein cores.  The results were obtained 
by simulating $100$ jammed packings of $N_R=24$ residues with 
amino acid frequencies that match protein cores. 
The probability distribution of packing fractions of protein cores is shown by the solid black line. Panels (a) and (b) 
are reprinted  with permission from 
[J. C. Gaines, W. W. Smith, L. Regan, and C. S. O'Hern, Phys. Rev. E, 
93, 032415, 2016.] Copyright (2016) by the American Physical Society.}
\label{phi_sim}
\end{figure}

As shown in Fig.~\ref{phi_sim} (a), the average packing fraction of
protein cores is $\phi_c \approx 0.56$~\cite{Gaines_2016}. This value
is much closer to packing fractions obtained for jammed packings of
frictional or elongated particles rather than $\phi_c = 0.71$-$0.74$
for packings with significant FCC crystalline order as proposed in
earlier studies~\cite{Richards_1974, Liang_2001, Tsai_1999}.  (See
Section \ref{packing_section}.) The most significant difference
between the recent and prior studies is the use of a well-calibrated
explicit hydrogen model instead of an extended atom model.

To assess the effect of backbone connectivity on the packing
efficiency in protein cores, we performed discrete element simulations
to compress amino acid monomers into static ({\it i.e.} force and
torque-balanced) jammed packings. (See the Appendix for a more
detailed description of the packing-generation protocol.) We
initialized the system by randomly inserting $N_R$ residues into a
cubic box (with periodic boundary conditions). We assumed that the
residues, which are composed of rigidly connected spherical atoms of
different sizes, interact via purely repulsive linear spring
forces. We then compress the system by small packing fraction
increments $\Delta \phi$, followed by energy minimization.  For
sufficiently small $\Delta \phi$, the form of the purely repulsive
potential does not influence the structure of the final packings. For
jammed packings, the total potential energy per residue $U/N_R > 0$
following energy minimization.  In contrast, unjammed packings will
possess $U/N_R = 0$ after energy minimization. In this case, atomic
motions can occur in the system without a concomitant increase in the
total potential energy. Thus, we can identify the packing fraction at
jamming onset $\phi_J$ as the one at which the minimized total
potential increases above a small threshold~\cite{Gao_2006}.

We studied mixtures of $N_R$ residues with the fractions of Ala, Ile,
Leu, Met, Phe, and Val residues matching the percentages found in
protein cores.  (We focused on non-polar residues, but because Gly has
no side chain and Cys can form disulfide bonds, these were not
included in the simulations.)  These simulations generate disordered
jammed packings with $ \phi = 0.56$ similar to that found in protein
cores (Fig. \ref{phi_sim} (b)).  These results indicate that the
connectivity of the protein backbone does not impose significant
constraints on the free volume in protein cores.

\begin{wrapfigure}{R}{3in}
\vspace{-.05in}
\includegraphics[width = 3in]{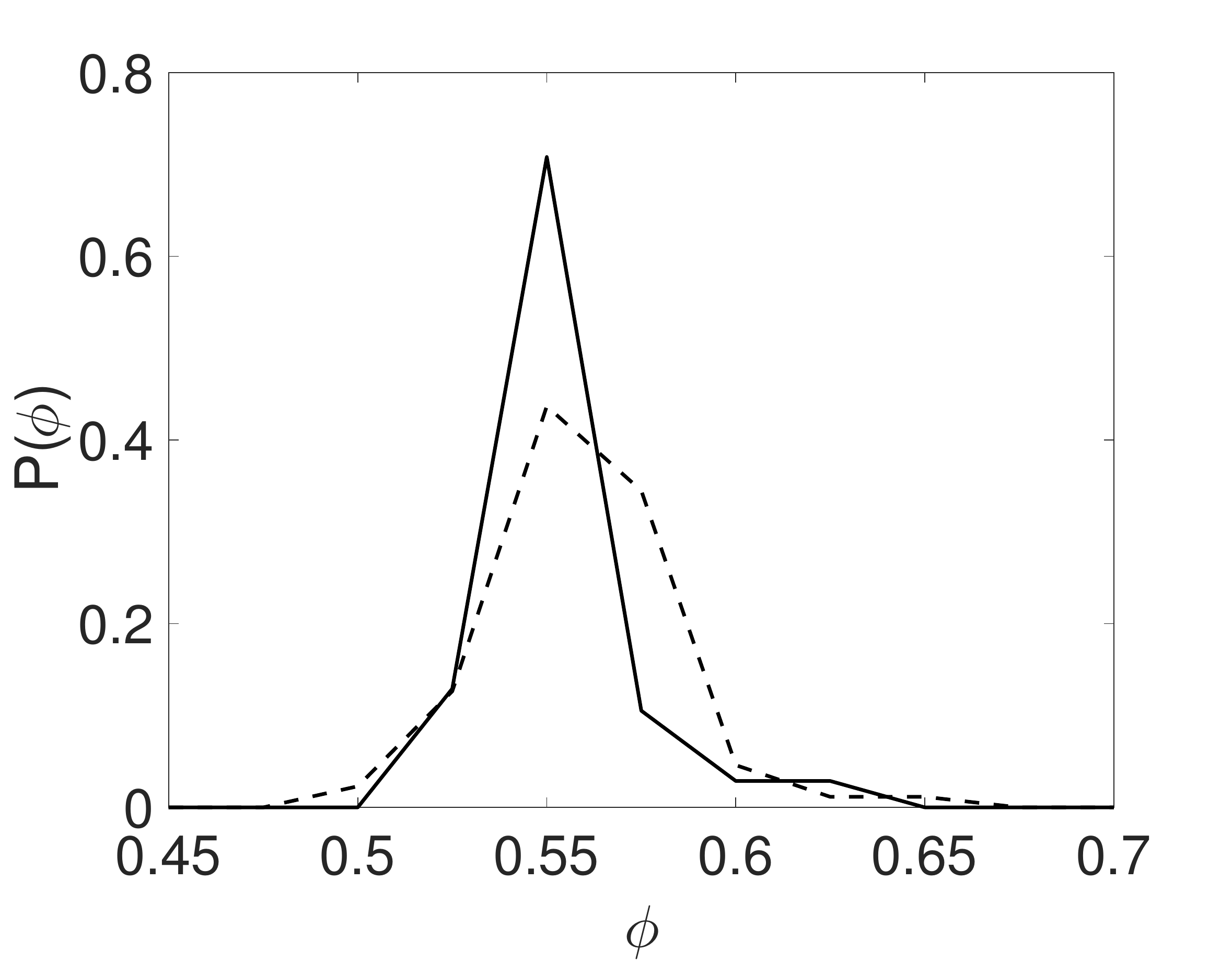}
\caption{The distribution of packing fractions $P(\phi)$ for core (solid line) and interface (dotted line) residues from high-resolution protein crystal structures.}
\label{interface}
\end{wrapfigure}

To further analyze the packing efficiency in protein cores, we also
calculated the distribution of the local packing fractions ({\it i.e.}
$\phi$ for each residue type) in protein cores for both protein
crystal structures and simulations.  We find that the distributions of
the local packing fractions for each residue type have similar average
values, differing by $< 5$\%.  In addition, the average values for the
local packing fractions are similar to the global average in the core
with standard deviations that are slightly larger, which reflects the
fact that the local packing fraction is obtained by averaging over
fewer atoms than the global packing fraction.  We also find that the
average packing fraction of each amino acid type is similar to the
average packing fraction in protein cores, except for Ala, which does
not have a side chain dihedral angle degree of freedom. The similarity
of the average packing fraction for individual residues and the
average packing fraction in protein cores suggests that there are only
small variations of the packing fraction within each protein core.

We also investigated the packing efficiency of protein-protein
interfaces.  To do this, we compiled a protein-interface database of
$123$ crystal structures containing protein-protein and
protein-peptide binding pairs. The structures are composed of both
homo- and heterodimers with resolution $\leq1.5$~\AA~and less than $50$\%
sequence identity. A core-interface residue is defined as any residue
that is a surface residue in the individual protein monomers, but is
completely buried after binding. Several studies have shown that the
properties of protein-protein interfaces are similar to those of
protein cores~\cite{Bordner_2005, Tsai_1997}. Our analyses of protein
cores and interfaces confirm this by showing that they possess a
similar distribution of amino acids ({\it i.e.} primarily hydrophobic
residues with few charged and polar residues). We find that 73\% and
68\% of the residues in protein cores and interfaces, respectively,
are hydrophobic with similar frequencies for each amino acid.  In 
addition, both
the distribution of core packing fractions and interface packing
fractions are peaked near $0.56$ as shown in
Fig.~\ref{interface}. This result demonstrates that protein-protein interfaces
are packed similarly to protein cores.

\section{Protein core repacking}
\label{core_repacking}

Computational protein core repacking allows investigation of the
uniqueness of the side chain conformation of residues in protein
cores.  Unique side chain conformations for core residues would imply
that protein cores are jammed with very little free volume for
rearrangements of side chains. There are two categories of protein
core repacking investigations: one starts with all possible sequences
and seeks to recover the wild type sequence \cite{Dantas_2007,
  Dobson_2006} and the other starts with the wild type sequence and
seeks to recover the observed combination of side chain dihedral
angles and determine if alternative combinations are possible. Here we
focus on the second, where the side chains of core residues are
removed simultaneously and all side chain dihedral angle combinations
of the starting sequence are sampled. The energy of each conformation
is evaluated, the optimal conformation is predicted, and then compared
to the observed structure.

\begin{figure}
\centering
\subfloat{\includegraphics[width =3.2in]{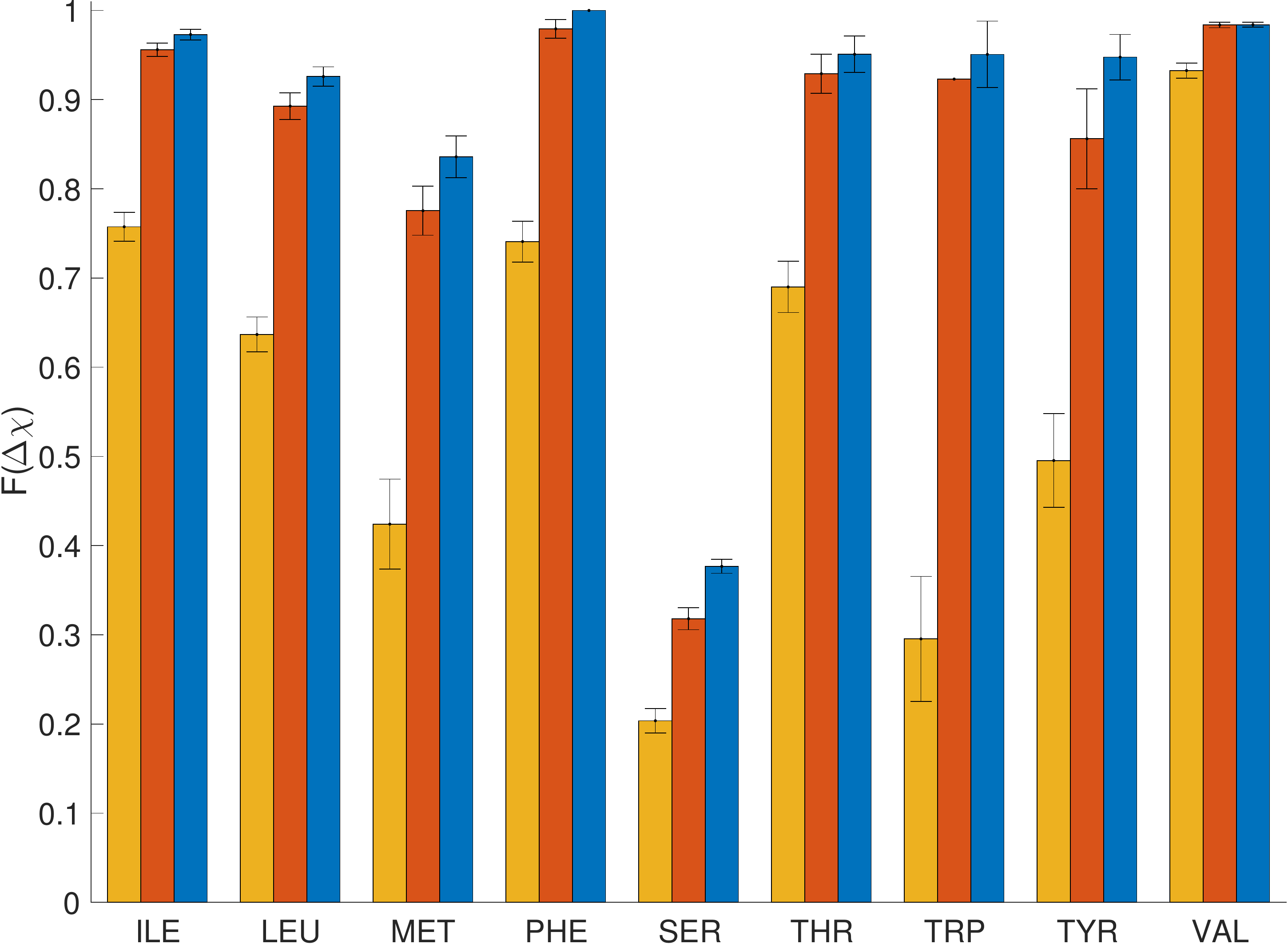}}
\subfloat{\includegraphics[width=3.2in]{Figure_4_repacking.pdf}}
\caption{(left) Single and (right) combined residue rotations in the context of the protein core: The fraction ($F(\Delta\chi))$ of each residue type for which the hard-sphere model prediction of the side chain conformation deviates by $\Delta\chi$ $<$ 10$\mathrm{^o}$ (yellow), 20$\mathrm{^o}$ (red), or 30$\mathrm{^o}$ (blue) from the crystal structure.}
\label{repacking}
\end{figure}

To study repacking of protein cores, we again use a hard-sphere plus
stereochemistry model. The cores of 221 proteins in the Dunbrack
Database \cite{Wang_2003, Wang_2005} were studied. As a way to model
the system at non-zero temperature and to improve the statistics,
variations in bond lengths and angles are implemented by replacing
each side chain with different instances of the side chain taken from
high-resolution protein crystal structures \cite{Dunbrack_1997}.  Core
residues were identified as described in Section \ref{phi_section}. As
described in previous work \cite{HS_Z2014, Gaines_2016}, the
hard-sphere model treats each atom $i$ as a sphere that interacts
pairwise with all other non-bonded atoms $j$ via the purely repulsive
Lennard-Jones potential:
\begin{equation}
\label{vrlj}
U_{\rm RLJ}(r_{ij}) = \frac{\epsilon}{72}\left[1-\left(\frac{\sigma_{ij}}{r_{ij}}\right)^6\right]^2 \Theta(\sigma_{ij} - r_{ij}),
\end{equation}
where $r_{ij}$ is the center-to-center separation between atoms $i$ and $j$, $\sigma_{ij} = (\sigma_i+\sigma_j)/2$, $\sigma_i/2$ is the radius of atom $i$, $\Theta(\sigma_{ij} - r_{ij})$ is the Heaviside step function, and $\epsilon$ is the strength of the repulsive interactions. Values for the atomic radii are listed in Section \ref{phi_section}.

Predictions of the side chain conformations of single amino acids are
obtained by rotating each of the side chain dihedral angles $\chi_1,
\chi_2, ..., \chi_n$ (with a fixed backbone conformation
\cite{Liu_2016}), and finding the lowest energy conformation of the
residue, where the total energy $U(\chi_1,...,\chi_n)$ includes both
intra- and inter-residue steric repulsive interactions. We then
calculate the Boltzmann weight of the lowest energy side chain
conformation of the residue, $P_i (\chi_1,....,\chi_n) \propto
e^{-U(\chi_1,...,\chi_n )/k_{B}T}$, where the small temperature,
$T/\epsilon$=$10^{-2}$, approximates hard-sphere-like interactions. We
select $50$ bond length and angle variants, and for each we find the lowest
energy dihedral angle conformation and corresponding $P_i
(\chi_1,....,\chi_n)$ values. We average $P_i$ over the variants to
obtain $P_m(\chi_1,....,\chi_n)$. We then compare the particular
dihedral angle combination $\{\chi_1^{HS},...,\chi_n^{HS}\} $
associated with the highest value of $P_m$ to the side chain of the
crystal structure $\{\chi_1^{xtal},...,\chi_n^{xtal}\}$. To assess the
accuracy of the hard-sphere model in predicting the side chain
dihedral angles of residues in protein cores, we calculate
\begin{equation}
\Delta \chi = \sqrt{(\chi_1^{xtal} - \chi_1^{HS})^2 +\ldots+(\chi_n^{xtal} - \chi_n^{HS})^2}.
\end{equation}
We determine the fraction $F(\Delta\chi)$ of residues of each type
with $\Delta\chi$ less than 10$\mathrm{^o}$, 20$\mathrm{^o}$, and
30$\mathrm{^o}$. (See Fig. \ref{repacking}.)

In Fig. \ref{repacking} (left), we investigate the accuracy of the
hard-sphere model in predicting the side chain dihedral angles of
single residues in protein cores. For each amino acid (Ile, Leu, Met,
Phe, Ser, Thr, Trp, Tyr, and Val), we calculate the fraction of
residues, $F(\Delta\chi)$, for which the predicted side chain dihedral
angle conformation is within 10$\mathrm{^o}$, 20$\mathrm{^o}$ and
30$\mathrm{^o}$ of the crystal structure value. Consistent with our
prior results, the hard-sphere model accurately predicts the side
chain dihedral angle combinations of single residues in the context of
the protein for Ile, Leu, Phe, Thr, Trp, Tyr, and Val ($\geq$ 90\%
within 30$\mathrm{^o}$) \cite{Caballero_2016}. This result emphasizes
that the purely repulsive hard-sphere model can accurately predict the
side chain dihedral angle combinations for nonpolar and uncharged
amino acids.

We find that the hard-sphere model is unable to predict with high
accuracy the observed side chain conformations for two residues that
we studied: Ser and Met. Our results for Met are consistent with those
found in Virrueta {et al.}~\cite{Virrueta_2016}. In this prior work,
we found that local steric interactions were insufficient to predict
the shape of the $P(\chi_3)$ distribution for Met.  It was necessary
to add attractive atomic interactions to the hard-sphere model to
reproduce the observed $P(\chi_3)$. Here, using only repulsive
interactions, we predict $\approx$ 80\% of Met residues are within
30$\mathrm{^o}$ of the crystal structure.  Our results for Ser (only
38\% within 30$\mathrm{^o}$) are also consistent with our prior work
in Caballero {\it et al.}~\cite{Caballero_2016}. We speculate that because
the side chain of Ser is small, hydrogen-bonding interactions must be
included to correctly place its side chain. In contrast, we suggest
that the more bulky Thr and Tyr side chains cause steric interactions
to determine the positioning of their side chains, even though they
are able to form hydrogen bonds \cite{HS_Z2012}.

\begin{SCfigure}[][t]
\includegraphics[width = 3in]{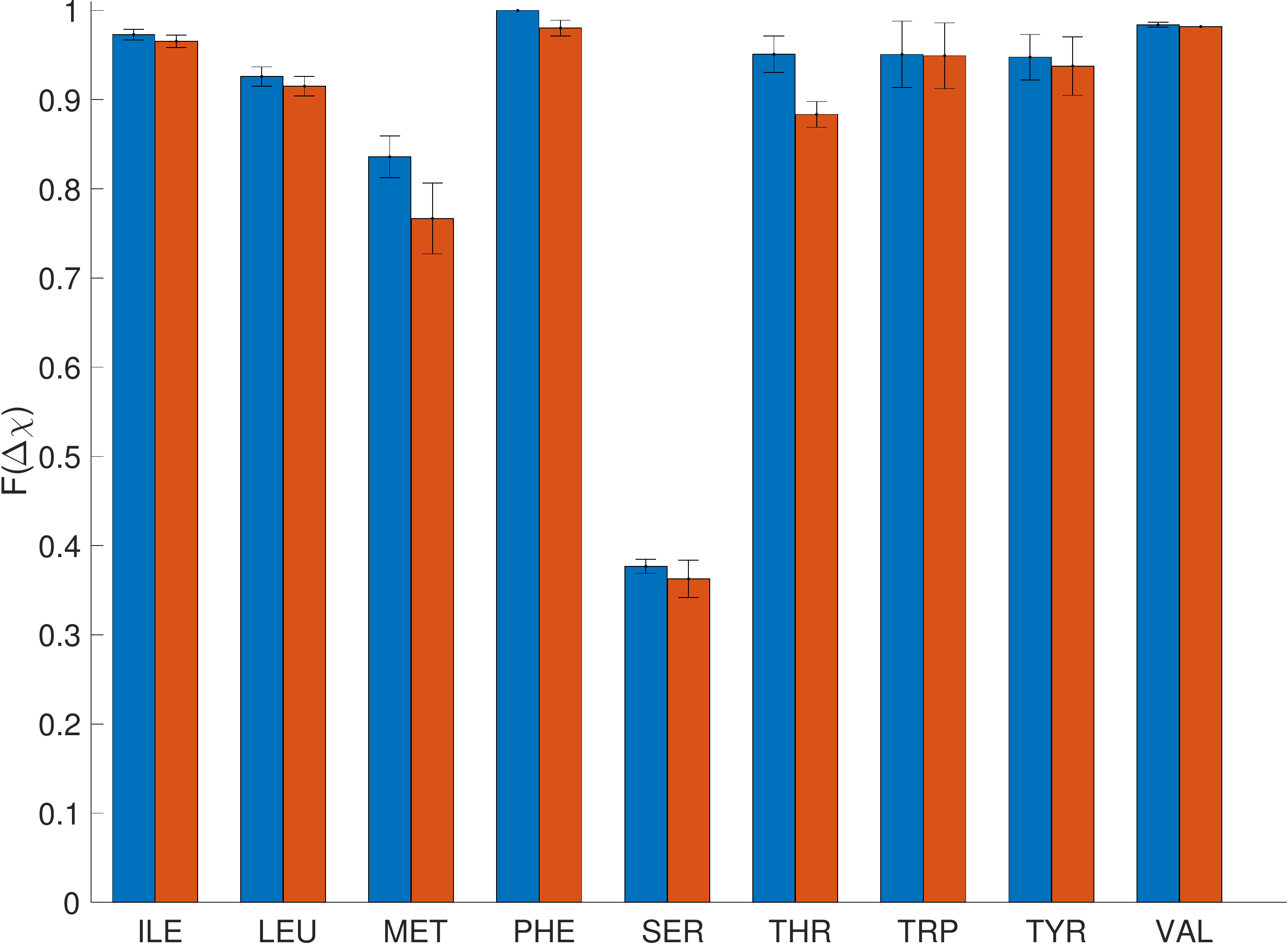}
\caption{Comparison of the accuracy of single and combined rotations
for core residues in 221 proteins \cite{Wang_2003, Wang_2005}. Each
bar shows the fraction of residues, $F(\Delta\chi)$, for which the
hard-sphere model prediction of the side chain conformation has
$\Delta \chi < 30\mathrm{^o}$ for single (blue) or combined (red)
rotations.}
\label{repacking_compare}
\end{SCfigure}

In addition to single residue rotations, we performed core repacking
using combined rotations of interacting core residues in each
protein~\cite{Gaines_2017}. For the combined rotation method, all
residues in an interacting cluster are rotated simultaneously (with
fixed backbone conformations), and the global minimum energy
conformation is identified.  A cluster of interacting residues is
defined such that side chain atoms of each residue in the cluster
interact with one or more other residues in the cluster, but do not
interact with the side chains of other core residues in the protein.

Single and combined rotations have the same prediction accuracy
(Figs. \ref{repacking} and \ref{repacking_compare}), which shows that
there are very few arrangements of the residues in a protein core that
are sterically allowed and that the side chain conformations of most
core residues are dominated by packing constraints. This result
implies that there are no alternative sterically allowed conformations
of core residues other than those in the crystal structure.  If
alternative sterically allowed conformations existed, we would have
found them using the collective repacking method and thus the
prediction accuracy would have dramatically decreased relative to the
value for single residue rotations.  It does not.  Thus, the results for collective
repacking reveal that the structures of protein cores are uniquely
specified by steric interactions. This conclusion is consistent with
those reached by Word {\it et al.}~\cite{Word_1999}, where they found
that ``in a well-packed core region, it is rare that a bond angle can
be rotated much in either direction without producing clashes.''

\section{Jammed packings of spherical and nonspherical particles}
\label{packing_section}

\begin{wrapfigure}{R}{3in}
\includegraphics[width = 3in]{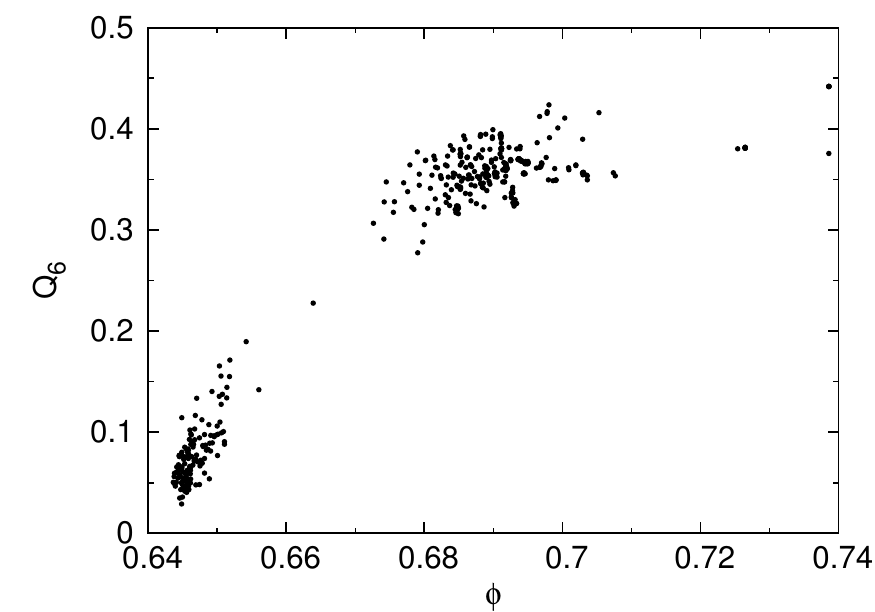}
\caption{Global bond orientational order parameter $Q_6$ versus packing 
fraction $\phi$ for $100$ jammed packings of monodisperse spheres.}
\vspace{-.25in}
\label{Q6}
\end{wrapfigure}

A strict definition of jamming means that a disordered system is
solid-like and possesses a static shear modulus~\cite{Ohern_2003}.
However, jamming also implies that a system is confined to a small
region of configuration space, such that little or no motion of the
constituent particles can occur.  The results presented in
Secs.~\ref{phi_section} and~\ref{core_repacking} provide several
indications that residues in protein cores are jammed in this latter
sense.  First, for nearly all protein cores, single and collective
repacking give the same side chain dihedral angle combinations found
in the protein crystal structures.  This result emphasizes that there
are no alternative low energy conformations for core residues. Second,
the packing fraction of protein cores is $\approx 0.56$, which is
similar to those reported for disordered jammed packings of
frictional~\cite{Silbert_2010} and elongated
particles~\cite{Zhao_2012,Donev_2007,Schreck_2012}.

In this section, we present the results of simulations of jammed
packings in three spatial dimensions (3D) for a wide variety of
particle shapes including monodisperse spheres, polydisperse spheres,
spheres with varying sizes of asperities (or ``bumps''), ellipsoids,
ellipsoids with varying sizes of asperities, and non-axisymmetric,
elongated particles.  This range of shapes allows us to study the
influence of the particle aspect ratio and surface bumpiness on the packing
fraction and determine which particle shapes produce packing fractions
that match the packing fraction of residues in protein cores.

\begin{figure}[!t]
\centering
\includegraphics[width = 4in]{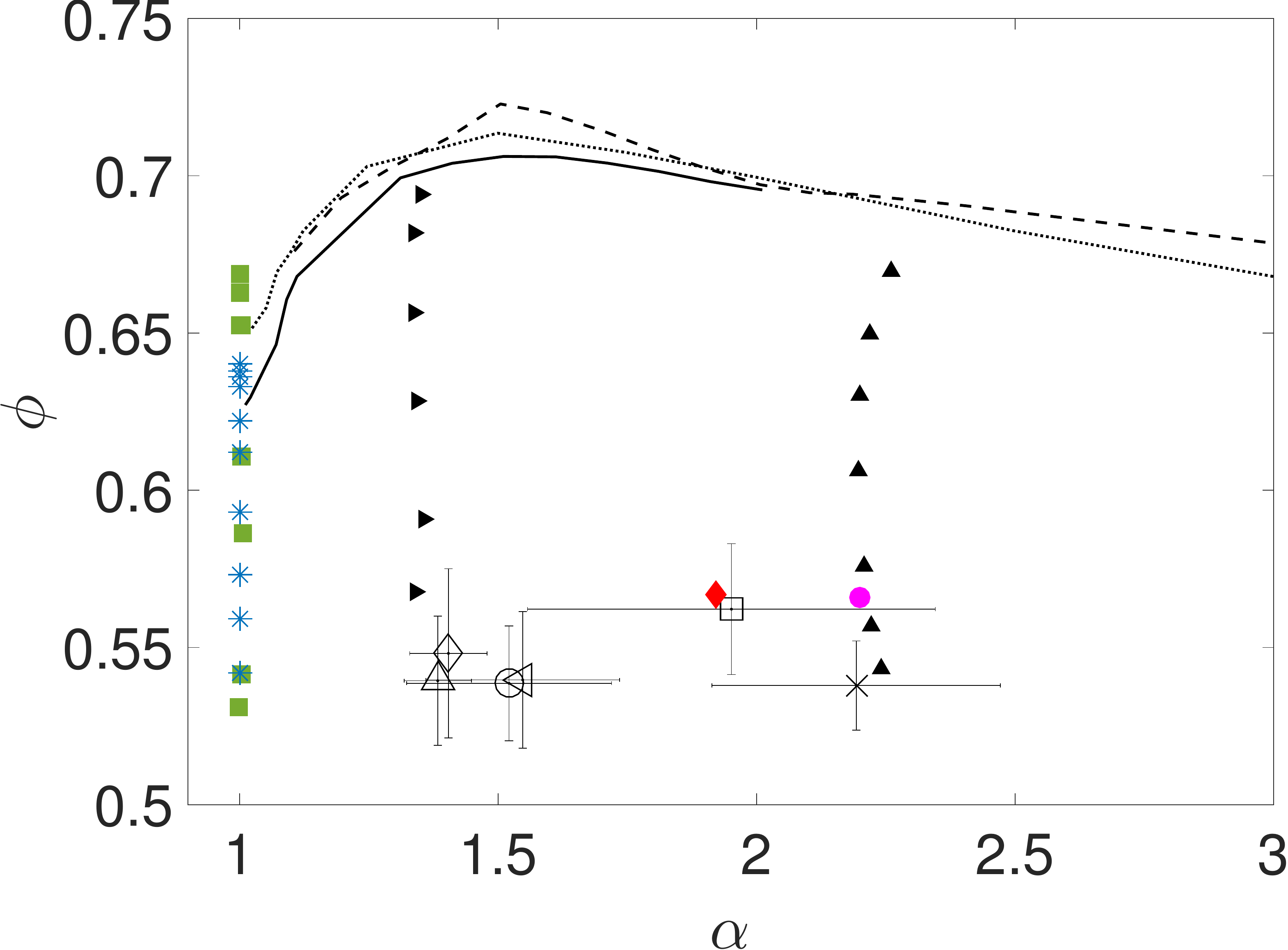}
\caption{Jammed packing fraction $\phi$ versus aspect ratio $\alpha$
for frictional spheres (blue asterisks) from Ref.~\cite{Silbert_2010}, 
bumpy (green triangles) spheres,
smooth, prolate ellipsoids of revolution from Refs.~\cite{Donev_2007} (dotted line)
and~\cite{Schreck_2012} (solid line) and spherocylinders (dashed
line) from Ref.~\cite{Zhao_2012}. The static friction coefficient for the
frictional spheres varies from $\mu = 10^{-4}$ to $10$ from top to
bottom.  For the bumpy spheres (Fig.~\ref{Shape_examples} (a) and
(b)), twelve bumps are placed on the vertices of an icosohedron, and
the relative sizes of the bumps are decreased to increase the bumpiness
$B$ from $\approx 10^{-2}$ to $0.15$ from top to bottom.
We also show the packing fraction and aspect ratio for 
Ala (open diamond), Ile (open leftward triangle), Leu (open circle), 
Met (open square),
Phe (x), and Val (open upward triangle) residues in protein cores. The error 
bars indicate the root-mean-square fluctuations from averaging over 
instances of each residue with different backbone and side chain 
conformations.  Results for bumpy ellipsoids are indicated by 
the filled rightward and upward triangles and results for the 
non-axisymmetric shapes in
Fig.~\ref{Shape_examples} (g) and (h) are indicated by the filled diamond and 
circle, respectively.}
\label{aspect_phi}
\end{figure}

We start the discussion with jammed packings of monodisperse
spheres. In monodisperse systems, the packing fraction depends on the
degree of order that is present in the system.  For example, in
Fig.~\ref{Q6}, we show that the packing fraction varies with the
global bond orientational order parameter
$Q_6$~\cite{Jin_2010,Truskett_2000}, which measures the degree to
which the separation vectors connecting a given particle and its
nearest neighbors are consistent with icosohedral symmetry. $Q_6
\approx 0.57$ for perfect FCC crystalline sphere packings with $\phi
\approx 0.74$.  The packing fraction for jammed packings of
monodisperse spheres decreases as $Q_6$ decreases, reaching random
close packing $\phi \approx 0.64$ in the limit $Q_6 \rightarrow
0$~\cite{Zhang_2014}. Jammed packings with different values of $Q_6$
can be obtained by varying the rate at which kinetic energy is
drained from the system~\cite{Ashwin_2012}.  For the present studies,
we consider the limit of fast quenching rates, which gives rise
to disordered packings.

Particle size differences can strongly decrease a system's tendency to
order.  In previous studies, we focused on jammed packings of bidisperse
spheres with half large spheres, half small spheres, and a modest
diameter ratio of $d=1.4$~\cite{Xu_2005,Gao_2006}.  It is difficult to
generate ordered packings of such bidisperse spheres using the
packing-generation methods employed here.  However, large size
ratios ($d \gtrsim 2.4$) can also increase the packing fraction
of jammed packings of polydisperse spheres.  In this case, small
spheres can fill in the gaps between contacting larger spheres.  For
example, Apollonian sphere packings~\cite{Farr_2010} characterized by
a continuous distribution of particle sizes possess packing fractions 
that approach $1$.

In the hard-sphere model of proteins, we consider seven atom types
with differing diameters.  The largest diameter ratio is $d=1.8$
between sulfur (which is rare) and hydrogen atoms; the next largest
diameter ratio ($d=1.5$) is between sp$^3$ carbon and hydrogen
atoms. Thus, we expect that jammed sphere packings composed of mixtures of
atoms with the same sizes and number fractions as in protein cores
will have packing fraction $\phi \approx 0.64$.  This result was shown
previously in Ref.~\cite{Gaines_2016}.  Thus, jammed packings composed
of individual spheres with polydispersity that matches atom size
differences in protein cores possess packing fractions that are larger
than the values we observe in protein cores (Sec.~\ref{phi_section}).

We now consider jammed packings of symmetric elongated particles, {\it
  i.e.} spherocylinders and ellipsoids, as a function of the aspect
ratio $\alpha$. In Fig.~\ref{aspect_phi}, we show that the packing fraction
$\phi(\alpha)$ is qualitatively the same for jammed packings of
spherocylinders and ellipsoids. $\phi \approx 0.64$ for spherical
particles with $\alpha=1$, increases for $\alpha>1$, reaches a
peak near $\alpha \approx 1.5$ with $\phi > 0.7$, and then decreases to a
plateau value of $\phi \approx 0.68$ at large $\alpha$.

To compare the results for jammed packings of symmetric, elongated
particles to packings of amino acids presented in Sec.~\ref{phi_section}, we
define a generalized aspect ratio and surface bumpiness to characterize
the shape of composite particles made from collections of 
spheres. We define bumpiness by 
\begin{equation}
B = \sqrt{ \bigint d{\hat u}  \frac{\left( {\vec R}({\hat u}) - {\vec {\cal R}}
({\hat u}) \right)^2}{{\cal R}^2({\hat u})}},
\label{bumpiness}
\end{equation}
where ${\hat u}$ is a unit vector with an origin at the geometric
center of the composite particle, the integral is over all orientations 
of ${\hat u}$, ${\vec R}({\hat u})$ gives the
location on the surface of the composite particle along ${\hat u}$,
and ${\vec {\cal R}}({\hat u})$ gives the location on the surface of a
reference prolate ellipsoid of revolution along ${\hat u}$. The
bumpiness $B$ for a given composite particle will depend on the
orientation of the reference prolate ellipsoid axis ${\hat e}$ and the values of the
major $a$ and minor $b$ axes.

To define the aspect ratio $\alpha$ for composite particles, we find the 
reference prolate ellipsoid of revolution that yields the smallest 
bumpiness.  We first fix the reference ellipsoid axis ${\hat e}$ to be in the direction that gives  
the largest distance between the geometric center and the surface of 
the composite particle.  We then minimize $B({\hat e},a,b)$ over $a$ and $b$
at fixed ${\hat e}$, and define $\alpha =a/b$ for the optimal values of 
the major and minor axes of the reference ellipsoid.

Fig.~\ref{aspect_phi} shows the packing fraction versus aspect ratio
for Ala, Val, Ile, Leu, Met, and Phe residues in protein cores. As
discussed in Sec.~\ref{phi_section}, most core residues have packing
fractions near $0.55$-$0.56$.   The
aspect ratios of amino acids depend on the amino acid type and their
backbone and side chain conformations.  The average aspect ratios vary
from $\alpha \approx 1.4$ for Val to $\approx 2.3$ for Phe.  The error
bars in both $\phi$ and $\alpha$ are obtained from the
root-mean-square fluctuations over different instances of each residue
in protein cores. 

\begin{figure}[!t]
\includegraphics[width = 6.5in]{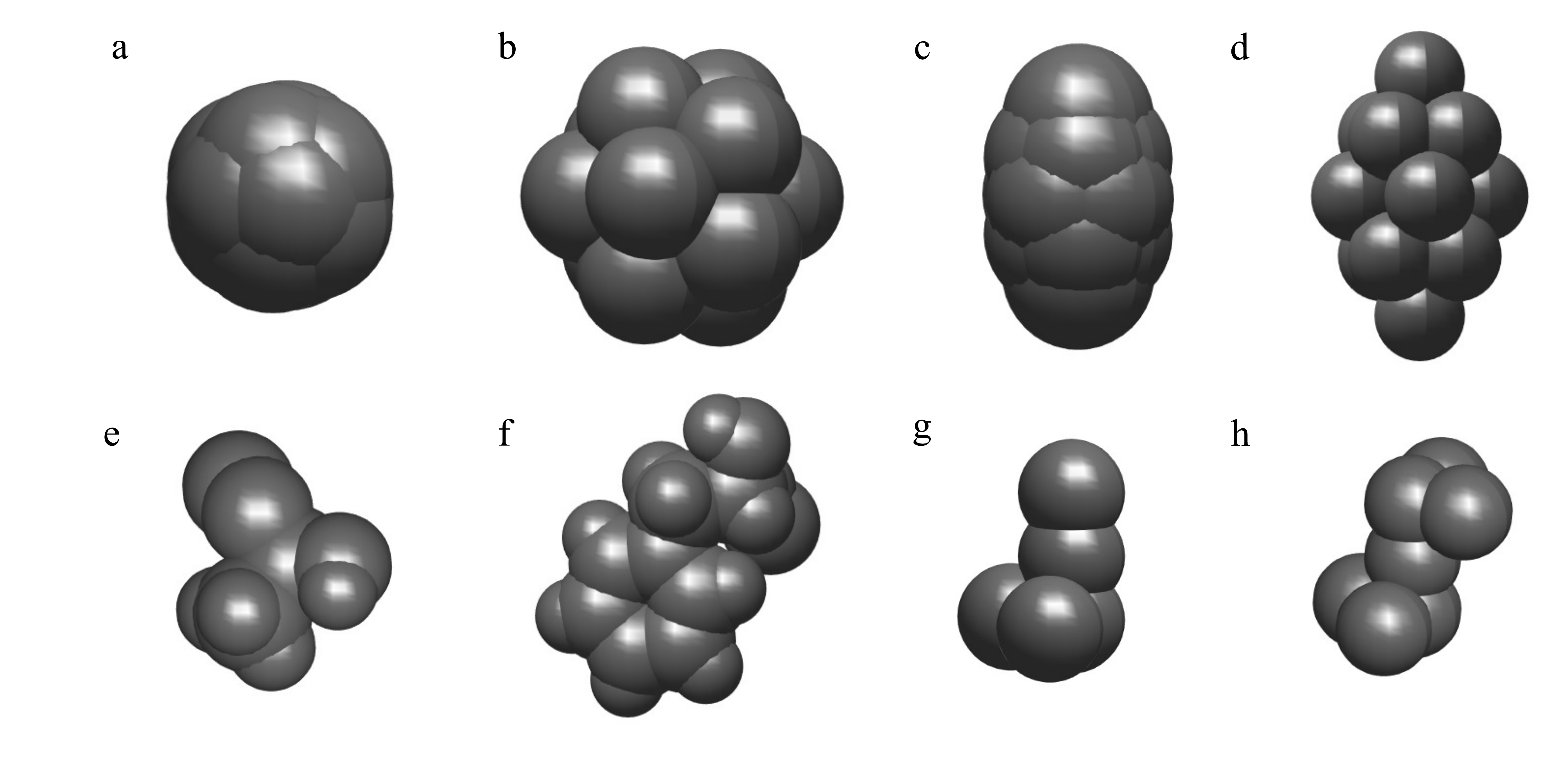}
\caption{Examples of the composite particle shapes investigated in the packing simulations: bumpy spheres with (a) $B=0.008$, $\alpha=1.00$ and (b) $B=0.113$, 
$\alpha=1.00$; bumpy ellipsoids with (c) $B=0.015$, $\alpha = 1.40$ and (e) $B=0.162$, 
$\alpha=1.40$; (e) Ala and (f) Phe residues; and (g,h) two examples of 
non-axisymmetric composite particles.}
\label{Shape_examples}
\end{figure}

The packing fraction $\phi \approx 0.55$-$0.56$ observed for amino
acids in protein cores with nominal aspect ratios in the range $1.4
\lesssim \alpha \lesssim 2.3$ is not consistent with the packing
fraction $\phi \approx 0.7$ obtained for jammed packings of ellipsoids
and spherocylinders with aspects ratios in the same range.  Thus,
elongated, smooth, axisymmetric particles are not sufficient to model
packings of amino acids in protein cores.

A method for decreasing the packing efficiency of particle packings is
to include frictional forces between particles or add asperities (or
``bumps'') to the surface of the particles as shown in
Fig.~\ref{Shape_examples} (a) and (b).  In prior work, we showed in 2D
that we could decrease the packing fraction of bidisperse disks from
random close to random loose packing (corresponding to more than a $10$\%
decrease in packing fraction) by increasing the bumpiness or effective
friction coefficient between disks~\cite{Papanikolaou_2013}.  In
Fig.~\ref{aspect_phi}, we include results from
Ref.~\cite{Silbert_2010} showing that the packing fraction of
frictional spheres (asterisks) in 3D decreases by a similar percentage
from $\phi \approx 0.64$ to $\approx 0.55$ as the static friction
coefficient $\mu$ increases from $10^{-4}$ to
$10$.

\begin{figure}[!b]
\centering
\includegraphics[width = 4in]{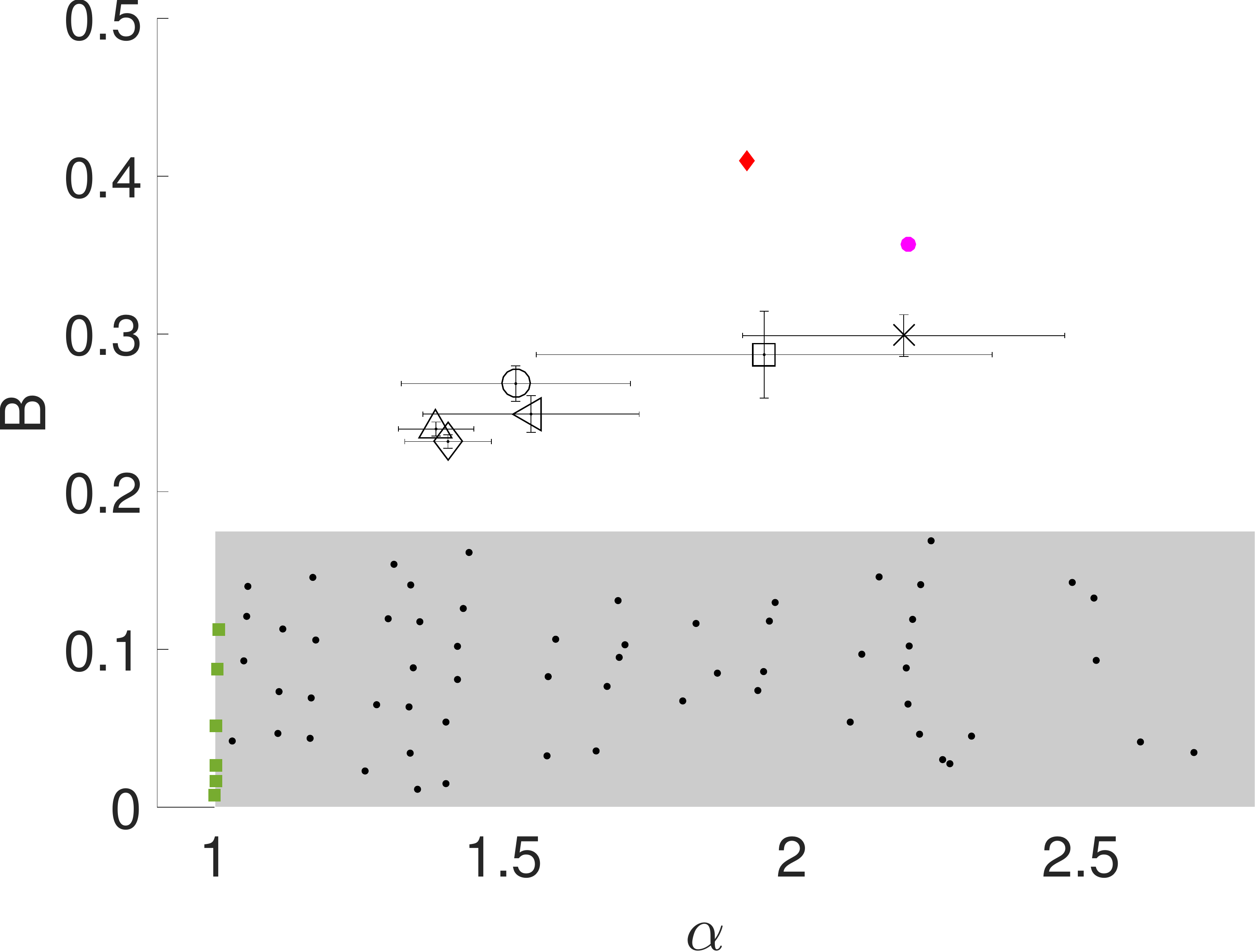}
\caption{Surface bumpiness $B$ versus aspect ratio $\alpha$ for
several particle shapes considered in the packing simulations. 
For bumpy spheres (green squares) with $\alpha = 1$ created by 
placing spheres on the vertices of an icosohedron, bumpiness can 
be varied over the range $0 \lesssim B \lesssim 0.15$. For 
prolate ellipsoids (black dots) with $8$ or $14$ spherical bumps 
(black dots), we can achieve maximum bumpiness values $B \approx 0.17$
over a wide range of $\alpha$ indicated by the grey rectangle.  We 
also show bumpiness versus aspect ratio for Ala
(diamond), Ile (leftward triangle), Leu (circle), Met (square), Phe
(x), and Val (upward triangle) residues in protein cores. $B$ 
and $\alpha$ for the non-axisymmetric particles in Fig.~\ref{Shape_examples}
(g) and (h) are given by the red diamond and magenta circle, respectively.}
\label{aspect_bumpy}
\end{figure}

We find similar results for bumpy spheres (green squares) in
Fig.~\ref{aspect_phi}. Here, the bumpy spheres are composite particles
made from twelve spheres arranged on the vertices of an icosohedron.
We decrease the ratio $r$ of the size of each sphere to the size of
the icosohedron to increase the bumpiness $B$.  We show in
Fig.~\ref{aspect_bumpy} that for bumpy spheres formed from an
icosohedron, we can generate $0 \lesssim B \lesssim 0.15$ (corresponding to $5
\gtrsim r \gtrsim 0.63$), which accounts for the decrease in packing
fraction of the green squares in Fig.~\ref{aspect_phi} from top to bottom.

As discussed above, amino acids cannot be modeled using spherical
shapes with $\alpha \approx 1$ or using elongated, smooth particles.
Thus, we performed studies of bumpy ellipsoids with $\alpha > 1$ to
model packings of amino acids in protein cores. For bumpy ellipsoids,
we place spheres on the surface of a reference prolate ellipsoid with
specified major and minor axes.  Two spheres were placed on the ends
of the reference ellipsoid and either $3$ or $4$ spheres were placed
at equal angular intervals on the ellipsoid surface at distances along
the long axis that divide the long axis into $3$ or $4$ equal
segments.  Thus, the bumpy ellipsoids we studied were made up of
either $8$ or $14$ spheres as shown in Fig.~\ref{Shape_examples} (c)
and (d).  In Fig.~\ref{aspect_bumpy}, we show that we can study
bumpiness values $B \lesssim 0.17$ over a wide range of aspect ratios
using this method for constructing bumpy axisymmetric elongated
particles.

In Fig.~\ref{aspect_phi}, we show the packing fraction for
jammed packings of bumpy ellipsoids over a range of bumpiness 
values for two aspect ratios, $\alpha \approx 
1.4$ and $2.25$, which spans the range of aspect ratios calculated for amino 
acids in protein cores.  For both aspect ratios, the packing fraction 
decreases from the values obtained from packings of smooth elongated 
particles to $\phi \approx 0.55$ as the bumpiness is increased from  
$B = 0.01$ to $0.17$.

An interesting point to note, as shown in Fig.~\ref{aspect_bumpy}, is
that amino acids found in protein cores ({\it e.g.} Ala and Phe in
Fig.~\ref{Shape_examples} (e) and (f)) possess bumpiness values
between $B=0.25$ and $0.3$, whereas bumpy axisymmetric shapes have $B
\lesssim 0.17$.  Thus, we also studied jammed packings composed of the
non-axisymmetric composite particles in Fig.~\ref{Shape_examples} (g)
and (h). Five spheres make up the composite particle pictured in panel
(g). Three are arranged in a straight line, and the other two spheres
are placed in a plane perpendicular to the long axis of the composite
particle and at an angular separation of $90^{\circ}$. The composite
particle in panel (h) contains $7$ spheres with two spheres each
placed at the top and bottom of the particle in planes perpendicular
to the long axis and in staggered orientations. The bumpiness and
aspect ratio of these non-axisymmetric composite particles is varied
by changing the size of the bumps compared to the size of the sphere
that circumscribes the composite particle.  For these two types of
non-axisymmetric particles, we were able to increase the maximum
bumpiness to $B \approx 0.4$, which is even larger than that of any
of the core amino acids (Fig.~\ref{aspect_bumpy}).
 
As shown in Fig.~\ref{aspect_phi}, the packing fractions for jammed
packings of the non-axisymmetric particles in
Fig.~\ref{Shape_examples} (g) and (h) (with $B = 0.33$ and $0.39$) are
$\phi \approx 0.56$. These results show that jammed packings of particles
with the same $B$ and $\alpha$ as those found for amino acids
yield the same packing fraction as amino acids in protein cores.

\section{Mutations in protein cores}
\label{mutations}

Additional insight into the packing efficiency in protein cores can be
obtained by examining the results from experimental studies of protein
core mutations. Several groups have experimentally investigated the
potential plasticity of protein cores by performing mutations, {\it
  i.e.} by changing the identities core amino acids. Lim and Sauer
simultaneously mutated several hydrophobic residues in the core of a
small protein, and used a genetic screen to identify those that were
functional and stable. They found that very few combinations of amino
acids other than the wildtype set resulted in a stable, folded protein
\cite{Lim_1989}. The functional new cores were dominated by
hydrophobic amino acids and the total side chain volumes were within
10\% of the original core volume. Combinations of residues outside of
these requirements were nonfunctional. Moreover, stereochemical
constraints further restricted the allowed sequence space. For
example, although many permutations of core residues can maintain the same
total volume and hydrophobicity in the core, they do not result in a
protein with the same structure and stability \cite{Lim_1989}. As a
result of hydrophobic, volume, and steric constraints, only 0.3\% of
60,000 sequences sampled are fully functional \cite{Lim_1989,
  Lim_1991}. These observations provide experimental support for the
dominance of steric interactions in protein cores. Similar
experimental results have been found in other proteins
\cite{Eriksson_1992, Eriksson_1993, Ishikawa_1993, Xu_1998}.

Liu, {\it et al.} investigated how mutations from small to large
residues in the core affect protein stability \cite{Liu_2000}. This
work illustrates the difficulty in generalizing the effects of a
particular type of mutation at different locations and in different
proteins. In this work, three Ala residues in the core of a small
protein were mutated, individually, to either Cys, Ile, Leu, Met, Phe,
Trp, or Val, and the resulting effect on protein stability was
determined.  They also solved the crystal structures of several of the
mutated proteins. They found that in all cases, to varying degrees, to
accommodate the larger amino acid side chain, the backbone
moved. Interestingly, at two of the three positions, even with
backbone movement, the protein with a larger side-chain was
destabilized relative to the protein with the original Ala.  However,
at one position, even large increases in volume (Ala to Phe or Trp)
could be accommodated by backbone movement to give a mutated protein
with similar stability to that of the parent protein.  Liu, {\it et
  al.}  hypothesized that this behavior was due to a cavity in the
protein near the mutation site, which allowed for more flexibility in
this region of the protein~\cite{Liu_2000}. (See also
Sec.~\ref{cavities_section}.)

This work shows that the protein core is not able to accommodate
mutations to larger residues without significant rearrangement and
subsequent destabilization of the original structure.  If substantial
empty space existed in the protein core, then mutations of this type
would likely have small effects because they would fill the existing
empty space and not require backbone rearrangements. Instead, backbone
rearrangements are necessary to accommodate larger amino acids,
supporting the idea that protein cores are tightly packed
\cite{Liu_2000}. This example also illustrates that much is still
unknown about protein core packing and how it controls protein
stability. The current state of knowledge is such that one can predict
neither the backbone movements in response to the incorporation of a
larger side chain, nor the changes in stability that result from these
structural changes.


\section{Conclusions and Future Directions} 
\label{cavities_section}

Our computational studies have established that protein cores are
comprised of irregularly shaped objects that are packed into
disordered jammed arrangements with $\phi \approx 0.56$~\cite{Gaines_2016}.  For a given
core, there are no alternative arrangements of the same amino acids
that are consistent with a well-packed core with no atomic
overlaps~\cite{Caballero_2016,Gaines_2017}. It has also been shown,
both experimentally and computationally, that there are a small number
of combinations of different core residues that can properly fit in
and fill a given core, and thus give rise to a stable folded protein
\cite{Lim_1989, Lim_1991, Liu_2000,  Dahiyat_1997, Kuhlman_2000}. There
are also experimental examples in which amino acids in the core are
substituted with ones that are either smaller or larger. Often such
substitutions result in changes in the backbone positions. With the
current state of understanding in the field, it is not possible to
reliably predict such movements. For some mutations, the rearranged
protein is as stable as the starting protein, for others it is less
stable. Again, the state of the art in computational modeling is such
that it is not possible to predict either the structure or the
stability of the repacked, rearranged protein.
 
Even dense packing of amino acids in protein cores results in some
void space not occupied by amino acids.  There has been some analysis
of voids in proteins using a range of probe
sizes~\cite{Cuff_2004,Liang_2001}. Various probe sizes are used to
identify void connectivity in the protein and to remove small
physically irrelevant voids. Obviously, an exceedingly small probe
({\it e.g.} radius $\lesssim 0.05$~\AA) will identify a large amount of void space,
because even the very smallest voids will be counted. Conversely, a
large probe ({\it e.g.} radius $\gtrsim 1.4$~\AA) will identify few, if
any, voids. A `reasonable' probe size to use seems to be around
$0.5$~\AA. Using such a probe size, Cuff, {\it et al.} examined void
statistics in a dataset of high-resolution protein
structures~\cite{Cuff_2004}. They found that the median total void
volume was $\approx 15 \mathrm{\AA}^3$ per residue. To put this into
perspective, a CH$_2$ group and a water molecule have a volume of
$\approx 25\mathrm{\AA}^3$, which indicates that the voids in protein
cores are small.  In future studies, we will consider the location and
size of buried voids to predict the consequences of changes of
amino acid size and sequence in protein cores.

\section{Acknowledgments}
\label{ack}
We gratefully acknowledge the support of the Raymond and Beverly
Sackler Institute for Biological, Physical, and Engineering Sciences
(to L.R., C.S.O., and J.C.G.), National Library of Medicine Training
Grant No. T15LM00705628 (to J.C.G.), and National Science Foundation
(NSF-PHY-1522467 to L.R., C.S.O. and J.C.G.). This work also benefited
from the facilities and staff of the Yale University Faculty of Arts
and Sciences High Performance Computing Center and the National
Science Foundation (Grant No. CNS-0821132), which in part funded
acquisition of the computational facilities.  We also thank Mark 
D. Shattuck for his input on measurements of void volumes in protein 
cores. 

\section{Appendix}
\label{appendix}

In this Appendix, we provide additional details that support the 
results presented in the main text. In 
Table~\ref{volume_res}, we provide the volume of the $11$ residues that 
occur most frequently in protein cores using the explicit hydrogen 
representation. Gly and Ala have the smallest volumes and Tyr and 
Trp have the largest. These values differ quantitatively from those 
obtained using the extended atom model. 

\begin{table}[h]
\begin{center}
\begin{tabular}{|c|c|}
	\hline
	Residue & Volume (\AA$^3$) \\ \hline
	Ala & 48.8\\ 
	Cys & 64.3 \\
	Gly & 35.6\\
	Ile & 88.1\\ 
	Leu & 88.1 \\
	Met & 92.7 \\
	Phe & 100.7 \\
	Thr & 69.0 \\
	Trp & 121.9\\
	Tyr & 107.5\\
	Val & 75.0 \\
	\hline
	
\end{tabular}
\caption{Volumes for the $11$ residues that occur most frequently in 
 protein cores using the explicit hydrogen representation.}
\label{volume_res}
\end{center}
\end{table}

We next describe the calculation of the error bars for the fraction
$F(\Delta \chi)$ of residues for which the prediction of the
hard-sphere model is less than $\Delta \chi$ from the observed side
chain conformation that are shown in Figs.~\ref{repacking} and
~\ref{repacking_compare}. To assess the accuracy of the hard-sphere
model in predicting the side chain dihedral angle conformations of
residues in protein cores, repacking calculations were performed using
$N_v = 300$ bond length and angle variants for each core residue.  For
each residue, we randomly select $M$ bond length and angle variants
out of the $N_v$ variants.  For each set of variants, we identified
the optimal side chain dihedral angle combination and calculated
$\Delta\chi$.  We then repeat this process $N$ times, which yields a
set of $N \Delta\chi$ values. We then calculated the mean fraction of
residues $F(\Delta \chi)$, which satisfy $\Delta \chi < 10^{\circ}$,
$20^{\circ}$, or $30^{\circ}$, and the standard deviation. We used
$N=50$ and $M=50$ for single residue rotations and $N=50$ and $M=30$
for combined rotations.

To understand how particle elongation and surface bumpiness affect
packing properties, we generated jammed packings of composite
particles formed from spheres. Each composite particle is composed of
$n$ spherical asperities placed on the vertices of an icosohedron or
locations on the surface of a prolate ellipsoid of
revolution. Spherical asperities $i$ and $j$ on composite particles
$C$ and $C$' interact via the pairwise potential $U^{CC'}_{ij} =
\frac{\epsilon}{2} (1-r_{ij}/\sigma_{ij})^2
\Theta(\sigma_{ij}-r_{ij})$, where $\epsilon$ is the energy scale of
the interaction, $r_{ij}$ is the distance between the centers of asperities $i$ and
$j$, $\sigma_{ij}=(\sigma_i + \sigma_j)/2$ is the average diameter
of asperities $i$ and $j$, and $\Theta$ is the Heaviside step
function. Thus, composite particles $C$ and $C$' interact via $U^{CC'} = \sum_{i,j} U^{CC'}_{ij}$.  The total 
potential energy of the system is $U = \sum_{C>C'} U^{CC'}$.

To find jammed packings, we employ a packing-generation protocol
similar to that in Ref.~\cite{Schreck_2012}. We first place $N$
composite particles randomly in a cubic periodic cell of unit size. At
each step we increase the asperity sizes $\sigma_i$ and bond lengths
$\delta_{ij}$ between asperities (fixing the ratios between $\sigma_i$ and
$\delta_{ij}$) corresponding to $\Delta \phi \approx 10^{-3}$, then we
relax the system to the nearest potential energy minimum using
dissipative dynamics, where the dissipative forces
are proportional to the composite particle velocities. If the
potential energy is zero after energy minimization ({\it i.e.} below a small
threshold $U/N < 10^{-4}$), we continue compressing; otherwise, we
decompress the system, where $\Delta \phi$ is halved each time we
switch from compression to decompression. We stop the packing-generation 
protocol when the potential
energy is nonzero and the average particle overlaps are
between $0.01$\% and $0.1$\%. We measure the final packing fraction at
jamming onset, which is insensitive to the choice of $\Delta \phi$
and the overlap threshold, provided they are 
sufficiently small.

\section{References}

\bibliographystyle{unsrt}
\bibliography{All_paper_Bib}

\begin{thebibliography}{10}

\bibitem{Dill_1990}
K.A. Dill.
\newblock Dominant forces in protein folding.
\newblock {\em Biochemistry}, 29:7133, 1990.

\bibitem{Rose_2006}
G.D. Rose, P.J. Fleming, J.R. Banavar, and A.~Maritan.
\newblock A backbone-based theory of protein folding.
\newblock {\em Proc. Natl. Acad. Sci. USA}, 103:16623, 2006.

\bibitem{Berman_2000}
H.M. Berman, J.~Westbrook, Z.~Feng, G.~Gilliland, T.N. Bhat, H.~Weissig, I.N.
  Shindyalov, and P.E. Bourne.
\newblock {The Protein Data Bank}.
\newblock {\em Nucleic Acids Res.}, 28:235, 2000.

\bibitem{Dunbrack_1997}
R.L. Dunbrack and F.E. Cohen.
\newblock Bayesian statistical analysis of protein side-chain rotamer
  preferences.
\newblock {\em Prot. Sci.}, 6:1661, 1997.

\bibitem{LoConte_1999}
L.~LoConte, C.~Chothia, and J.~Janin.
\newblock The atomic structure of protein-protein recognition sites.
\newblock {\em J. Mol. Biol.}, 285:2117, 1999.

\bibitem{Glaser_2001}
F.~Glaser, D.M. Steinberg, I.A. Vakser, and N.~Ben-Tal.
\newblock Residue frequencies and pairing preferences at protein-protein
  interfaces.
\newblock {\em Proteins: Struct., Funct., Bioinf.}, 43:89, 2001.

\bibitem{Keskin_2004}
O.~Keskin, C-J Tsa, H.~Wolfson, and R.~Nussinov.
\newblock A new, structurally nonredundant, diverse data set of protein-protein
  interfaces and its implications.
\newblock {\em Protein Sci.}, 13:1043, 2004.

\bibitem{Bordner_2005}
A.J. Bordner and R.~Abagyan.
\newblock Statistical analysis and prediction of protein-protein interfaces.
\newblock {\em Proteins}, 60:353, 2005.

\bibitem{Reichmann_2007}
D.~Reichmann, O.~Rahat, M.~Cohen, H.~Neuvirth, and G.~Schreiber.
\newblock The molecular architecture of protein-€"protein binding sites.
\newblock {\em Cur. Opin. Struc. Biol.}, 17(1):67, 2007.

\bibitem{Sheffler_2009}
W.~Sheffler and D.~Baker.
\newblock {RosettaHoles}: Rapid assessment of protein core packing for
  structure prediction, refinement, design and validation.
\newblock {\em Protein Sci.}, 18:229, 2009.

\bibitem{London_2010}
N.~London, D.~Movshovitz-Attias, and O.~Schueler-Furman.
\newblock The structural basis of peptide-protein binding strategies.
\newblock {\em Structure}, 18(2):188, 2010.

\bibitem{HS_Z2011}
A.Q. Zhou, C.S. O'Hern, and L.~Regan.
\newblock Revisiting the {Ramachandran} plot from a new angle.
\newblock {\em Protein Sci.}, 20:1166, 2011.

\bibitem{HS_Z2013}
A.Q. Zhou, D.~Caballero, C.S. O'Hern, and L.~Regan.
\newblock New insights into the interdependence between amino acid
  stereochemistry and protein structure.
\newblock {\em Biophys. J.}, 105:2403, 2013.

\bibitem{Gaines_2016}
J.C. Gaines, W.W. Smith, L.~Regan, and C.S. O'Hern.
\newblock Random close packing in protein cores.
\newblock {\em Physical Review E}, 93, 2016.

\bibitem{Engh_1991}
R.A. Engh and R.~Huber.
\newblock Accurate bond and angle parameters for {X}-ray protein structure
  refinement.
\newblock {\em Acta Crystallogr. A}, 47:392, 1991.

\bibitem{Allen_2002}
F.H. Allen.
\newblock The {Cambridge Structural Database}: {A} quarter of a million crystal
  structures and rising.
\newblock {\em Acta. Crystallogr. B}, 58:380, 2002.

\bibitem{Ramakrishnan_1963}
G.N. Ramachandran, C.~Ramakrishnan, and V.~Sasisekharan.
\newblock Stereochemistry of polypeptide chain configurations.
\newblock {\em J. Mol. Biol.}, page~95, 1963.

\bibitem{Ramakrishnan_1965}
C.~Ramakrishnan and G.~N. Ramachandran.
\newblock Stereochemical criteria for polypeptide and protein chain
  conformations.
\newblock {\em Biophys. J.}, 5:909, 1965.

\bibitem{Bryson_1995}
J.W. Bryson, S.F. Betz, H.S. Lu, D.J. Suich, H.X. Zhou, K.T. O'Neil, and W.F.
  DeGrado.
\newblock Protein design: A hierarchic approach.
\newblock {\em Science}, 270:935, 1995.

\bibitem{Munson_1996}
M.~Munson, S.~Balasubramanian, K.G. Fleming, A.D. Nagi, R.~O'Brien, J.M.
  Sturtevant, and L.~Regan.
\newblock What makes a protein a protein? {H}ydrophobic core designs that
  specify stability and structural properties.
\newblock {\em Protein Sci.}, 5:1584, 1996.

\bibitem{Smith_1995}
C.K. Smith and L.~Regan.
\newblock Guidelines for protein design: {T}he energetics of beta sheet side
  chain interactions.
\newblock {\em Science}, 270:980, 1995.

\bibitem{Richards_1974}
F.M. Richards.
\newblock The interpretation of protein structures: {T}otal volume, group
  volume distributions and packing density.
\newblock {\em J. Mol. Biol.}, 82:1, 1974.

\bibitem{Liang_2001}
J.~Liang and K.~Dill.
\newblock Are proteins well-packed?
\newblock {\em Biophys. J.}, 81:751, 2001.

\bibitem{Lim_1989}
W.A. Lim and R.T. Sauer.
\newblock Alternative packing arrangements in the hydrophobic core of lambda
  repressor.
\newblock {\em Nature}, 339:31, 1989.

\bibitem{Lim_1991}
W.A. Lim and R.T. Sauer.
\newblock The role of internal packing interactions in determining the
  structure and stability of a protein.
\newblock {\em J. Mol. Biol.}, 219(2):359, 1991.

\bibitem{Ohern_2003}
C.S. O'Hern, L.E. Silbert, A.J. Liu, and S.R. Nagel.
\newblock Jamming at zero temperature and zero applied stress: The epitome of
  disorder.
\newblock {\em Phys. Rev. E}, 68:011306, 2003.

\bibitem{Gekko_2015}
K.~Gekko.
\newblock {\em Volume and Compressibility of Proteins}, page~75.
\newblock Springer Netherlands, Dordrecht, 2015.

\bibitem{Chalikian_1995}
T.V. Chalikian, V.S. Gindikin, and K.J. Breslauer.
\newblock Volumetric characterizations of the native, molten globule and
  unfolded states of cytochromecat acidic {pH}.
\newblock {\em J. Mol. Biol.}, 250:291, 1995.

\bibitem{Gao_2015}
M.~Gao, H.~Zhou, and J.~Skolnick.
\newblock Insights into disease-associated mutations in the human proteome
  through protein structural analysis.
\newblock {\em Structure}, 23:1362, 2015.

\bibitem{Regan_2015}
L.~Regan, D.~Caballero, M.~R. Hinrichsen, A.~Virrueta, D.~M. Williams, and
  C.~S. O'Hern.
\newblock Protein design: Past, present, and future.
\newblock {\em Biopolymers Peptide Science}, 104:334, 2015.

\bibitem{Sheffler_2010}
W.~Sheffler and D.~Baker.
\newblock {RosettaHoles2}: A volumetric packing measure for protein structure
  refinement and validation.
\newblock {\em Protein Sci.}, 19:1991, 2010.

\bibitem{Wang_2003}
G.~Wang and R.L.~Dunbrack Jr.
\newblock {PISCES: A protein sequence culling server}.
\newblock {\em Bioinformatics}, 19:1589, 2003.

\bibitem{Wang_2005}
G.~Wang and R.L.~Dunbrack Jr.
\newblock {PISCES: Recent improvements to a PDB sequence culling server}.
\newblock {\em Nucleic Acids Res.}, 33:W94, 2005.

\bibitem{Tsai_1999}
J.~Tsai, R.~Taylor, C.~Chothia, and M.~Gerstein.
\newblock The packing density in proteins: {S}tandard radii and volumes.
\newblock {\em J. Mol. Biol.}, 290:253, 1999.

\bibitem{Word_1999}
J.M. Word, S.C. Lovell, J.S. Richardson, and D.C. Richardson.
\newblock Asparagine and glutamine: {U}sing hydrogen atom contacts in the
  choice of side-chain amide orientation.
\newblock {\em J. Mol. Biol.}, 285:1735, 1999.

\bibitem{Word_1999b}
J.M. Word, S.C. Lovell, J.S. Richardson, and D.C. Richardson.
\newblock Visualizing and quantifying molecular goodness-of-fit: {S}mall-probe
  contact dots with explicit hydrogen atoms.
\newblock {\em J. Mol. Biol.}, 285:1735, 1999.

\bibitem{HS_Z2012}
A.Q. Zhou, C.S. O'Hern, and L.~Regan.
\newblock The power of hard-sphere models: {Explaining side-chain dihedral
  angle distributions of Thr and Val}.
\newblock {\em Biophys. J.}, 102:2345, 2012.

\bibitem{HS_Z2014}
A.Q. Zhou, C.S. O'Hern, and L.~Regan.
\newblock Predicting the side-chain dihedral angle distributions of non-polar,
  aromatic, and polar amino acids using hard sphere models.
\newblock {\em Proteins}, 82:2574, 2014.

\bibitem{Bondi_1964}
A.~Bondi.
\newblock Vdw volumes and radii.
\newblock {\em J. Phys. Chem.}, 68:441, 1964.

\bibitem{CDC}
{Element data and radii, Cambridge Crystallographic Data Centre}.
\newblock http://www.ccdc.cam.ac.uk/products/csd/radii.
\newblock [Online; {A}ccessed December 4, 2011].

\bibitem{Seeliger_2007}
D.~Seeliger and B.L. {de Groot}.
\newblock {Atomic contacts in protein structures. {A} detailed analysis of
  atomic radii, packing, and overlaps}.
\newblock {\em Proteins Struct. Funct. Bioinf.}, 68:595, 2007.

\bibitem{Pauling_1948}
L.~Pauling.
\newblock {\em The Nature of the Chemical Bond}.
\newblock Cornell University Press, Ithaca, NY, 1948.

\bibitem{Porter_2011}
L.L. Porter and G.D. Rose.
\newblock {Redrawing the Ramachandran plot after inclusion of hydrogen-bonding
  constraints}.
\newblock {\em Proc. Natl. Acad. Sci. USA.}, 108:109, 2011.

\bibitem{Chothia_1975}
C.~Chothia.
\newblock Structural invariants in protein folding.
\newblock {\em Nature}, 254:304, 1975.

\bibitem{Li_1998}
A.J. Li and R.~Nussinov.
\newblock A set of van der {Waals }and coulombic radii of protein atoms for
  molecular and solvent-accessible surface calculation, packing evaluation, and
  docking.
\newblock {\em Proteins Struct. Funct. Bioinf.}, 32:111, 1998.

\bibitem{Momany_1974}
F.A. Mamony, L.M. Carruthers, and H.A. Scheraga.
\newblock Intermolecular potentials from crystal data. {III. D}etermination of
  empirical potentials and application to the packing configurations and
  lattice energies in crystals of hydrocarbons, carboxylic acids, amines, and
  amides.
\newblock {\em J. Phys. Chem.}, 78:1595, 1974.

\bibitem{Allinger_1980}
N.L. Allinger and Y.H. Yuh.
\newblock {\em Quantum Chemistry Program Exchange}, 12:395, 1980.

\bibitem{Rycroft_2009}
C.H. Rycroft.
\newblock {Voro++: A three-dimensional Voronoi cell library in C++}.
\newblock {\em Chaos}, 19:041111, 2009.

\bibitem{Caballero_2016}
D.~Caballero, A.~Virrueta, C.S. O'Hern, and L.~Regan.
\newblock Steric interactions determine side-chain conformation in protein
  cores.
\newblock {\em Protein Eng., Des. Sel.}, 29:367, 2016.

\bibitem{Gao_2006}
G.-J. Gao, J.~Blawzdziewicz, , and C.S. O'Hern.
\newblock Studies of the frequency distribution of mechanically stable disk
  packings.
\newblock {\em Phys. Rev. E}, 74:061304, 2006.

\bibitem{Tsai_1997}
C.J. Tsai, S.L. Lin, H.J. Wolfson, and R.~Nussinov.
\newblock Studies of protein-protein interfaces: A statistical analysis of the
  hydrophobic effect.
\newblock {\em Protein Sci.}, 6:53, 1997.

\bibitem{Dantas_2007}
G.~Dantas, C.~Corrent, S.L. Reichow, J.J. Havranek, Z.M. Eletr, N.G. Isern,
  B.~Kuhlman, G.~Varani, E.A. Merritt, and D.~Baker.
\newblock High-resolution structural and thermodynamic analysis of extreme
  stabilization of human procarboxypeptidase by computational protein design.
\newblock {\em J Mol. Biol.}, 366:1209, 2007.

\bibitem{Dobson_2006}
N.~Dobson, G.~Dantas, D.~Baker, and G.~Varani.
\newblock High-resolution structural validation of the computational redesign
  of human {U1A} protein.
\newblock {\em Structure}, 14:847, 2006.

\bibitem{Liu_2016}
H.~Liu and Q.~Chen.
\newblock Computational protein design for given backbone: recent progresses in
  general method-related aspects.
\newblock {\em Curr. Opin. Struct. Biol.}, 39:89, 2016.

\bibitem{Virrueta_2016}
A.~Virrueta, C.~S. O'Hern, and L.~Regan.
\newblock Understanding the physical basis for the side chain conformational
  preferences of met.
\newblock {\em Proteins: Struct., Funct., Bioinf.}, 84:900, 2016.

\bibitem{Gaines_2017}
J.~C. Gaines, A.~Virrueta, S.J. Fleishman, C.~S. O'Hern, and L.~Regan.
\newblock Collective repacking reveals that the structure of protein cores are
  uniquely specified by steric repulsive interactions.
\newblock {\em Protein Eng. Des. Sel.}, 2017.

\bibitem{Silbert_2010}
L.E. Silbert.
\newblock Jamming of frictional spheres and random loose packing.
\newblock {\em Soft Matter}, 6:2918, 2010.

\bibitem{Zhao_2012}
J.~Zhao, S.~Li, R.~Zou, and A.~Yu.
\newblock Dense random packings of spherocylinders.
\newblock {\em Soft Matter}, 8:1003, 2012.

\bibitem{Donev_2007}
A.~Donev, R.~Connelly, F.H. Stillinger, and S.~Torquato.
\newblock Underconstrained jammed packings of nonspherical hard particles:
  Ellipses and ellipsoids.
\newblock {\em Phys. Rev. E}, 75:051304, 2007.

\bibitem{Schreck_2012}
C.~F. Schreck, M.~Mailman, B.~Chakraborty, and C.~S. O'Hern.
\newblock Constraints and vibrations in static packings of ellipsoidal
  particles.
\newblock {\em Phys. Rev. E}, 85:061305, 2012.

\bibitem{Jin_2010}
Y.~Jin and H.~A. Makse.
\newblock A first-order phase transition defines the random close packing of
  hard spheres.
\newblock {\em Physica A}, 389:5362, 2010.

\bibitem{Truskett_2000}
T.~M. Truskett, S.~Torquato, and P.~G. Debenedetti.
\newblock Quantifying disorder in equilibrium and glassy sphere packings.
\newblock {\em Phys. Rev. E}, 62:993, 2000.

\bibitem{Zhang_2014}
K.~Zhang, W.W. Smith, M.~Wang, Y.~Liu, J.~Schroers, M.D. Shattuck, and C.S.
  O'Hern.
\newblock Connection between the packing efficiency of binary hard spheres and
  the glass-forming ability of bulk metallic glasses.
\newblock {\em Phys. Rev. E}, 90:032311, 2014.

\bibitem{Ashwin_2012}
S.~S. Ashwin, J.~Blwzdziewicz, C.~S. O'Hern, and M.~D. Shattuck.
\newblock Calculations of the basin volumes for mechanically stable packings.
\newblock {\em Phys. Rev. E}, 85:061307, 2012.

\bibitem{Xu_2005}
N.~Xu, J.~Blawzdziewicz, and C.S. O'Hern.
\newblock {Reexamination of random close packing: Ways to pack frictionless
  disks}.
\newblock {\em Phys. Rev. E}, 71:061306, 2005.

\bibitem{Farr_2010}
R.S. Farr and E.~Griffiths.
\newblock Estimate for the fractal dimension of the {A}pollonian gasket in d
  dimensions.
\newblock {\em Phys Rev E Stat Nonlin Soft Matter Phys}, 81, 2010.

\bibitem{Papanikolaou_2013}
S.~Papanikolaou, C.S. O'Hern, and M.D. Shattuck.
\newblock Isostaticity at frictional jamming.
\newblock {\em Phys. Rev. Lett.}, 110:198002, 2013.

\bibitem{Eriksson_1992}
A.E. Eriksson, W.A. Baase, X.J. Zhang, D.W. Heinz, M.~Blaber, E.P. Baldwin, and
  B.W. Matthews.
\newblock Response of a protein structure to cavity-creating mutations and its
  relation to the hydrophobic effect.
\newblock {\em Science}, 255:178, 1992.

\bibitem{Eriksson_1993}
A.E. Eriksson, W.A. Baase, and B.W. Matthews.
\newblock Similar hydrophobic replacements of {Leu99 and Phe153 } within the
  core of {T4} lysozyme have different structural and thermodynamic
  consequences.
\newblock {\em J. Mol. Biol.}, 229:747, 1993.

\bibitem{Ishikawa_1993}
K.~Ishikawa, H.~Nakamura, K.~Morikawa, and S.~Kanaya.
\newblock Stabilization of {Escherichia} coli ribonuclease {HI} by
  cavity-filling mutations within a hydrophobic core.
\newblock {\em Biochemistry}, 32:6171, 1993.

\bibitem{Xu_1998}
J.~Xu, W.A. Baase, E.~Baldwin, and B.W. Matthews.
\newblock The response of {T4} lysozyme to large-to-small substitutions within
  the core and its relation to the hydrophobic effect.
\newblock {\em Protein Sci.}, 7:158, 1998.

\bibitem{Liu_2000}
R.~Liu, W.A. Baase, and B.W. Matthews.
\newblock The introduction of strain and its effects on the structure and
  stability of {T4} lysozyme.
\newblock {\em J. Mol. Biol.}, 295:127, 2000.

\bibitem{Dahiyat_1997}
B.I. Dahiyat and S.L. Mayo.
\newblock Probing the role of packing specificity in protein design.
\newblock {\em Proc. Natl. Acad. Sci. U.S.A.}, 94(19):10172, 1997.

\bibitem{Kuhlman_2000}
B.~Kuhlman and D.~Baker.
\newblock Native protein sequences are close to optimal for their structures.
\newblock {\em Proc. Natl. Acad. Sci. USA.}, 97:10383, 2000.

\bibitem{Cuff_2004}
A.L. Cuff and A.C.R. Martin.
\newblock Analysis of void volumes in proteins and application to stability of
  the p53 tumour suppressor protein.
\newblock {\em J. Mol.Biol.}, 344:1199, 2004.

\end{thebibliography}

\end{document}